\documentclass[reprint,aps,prx,twocolumn,superscriptaddress,floatfix,nofootinbib]{revtex4-2}

\usepackage{amsmath,amsfonts,calc}
\usepackage[colorlinks, linkcolor = magenta, citecolor = cyan, bookmarksdepth=2, linktocpage=true, hypertexnames=true]{hyperref}
\usepackage[capitalize]{cleveref}
\usepackage{mathtools}
\usepackage{graphicx}
\usepackage[utf8]{inputenc}
\usepackage{braket}
\usepackage{gensymb}
\usepackage{soul}

\usepackage[left]{lineno}

\begin{document}

\title{Asymmetry Control in a Parametric Oscillator for the Quantum Simulation of Chemical Activation} 

\author{Alejandro Cros Carrillo de Albornoz}
\thanks{These authors contributed equally}
\email{a.croscarrillo@gmail.com}
\affiliation{Department of Applied Physics and Physics, Yale University, New Haven, CT 06520, USA}
\affiliation{Department of Physics and Astronomy, University College London, London WC1E 6BT, UK}

\author{Rodrigo G. Cortiñas}
\thanks{These authors contributed equally}
\altaffiliation{Present address: Google Quantum AI, Santa Barbara, CA, USA}
\email{rodrigogc@google.com}
\affiliation{Department of Applied Physics and Physics, Yale University, New Haven, CT 06520, USA}

\author{Max Schäfer}
\thanks{These authors contributed equally}
\altaffiliation{Present address: Department of Physics, University of California, Santa Barbara, CA, USA}
\email{maxschaefer@ucsb.edu}
\affiliation{Department of Applied Physics and Physics, Yale University, New Haven, CT 06520, USA}

\author{Nicholas E. Frattini}
\altaffiliation{Present address: Nord Quantique, Sherbrooke, QC J1J 2E2, Canada}
\affiliation{Department of Applied Physics and Physics, Yale University, New Haven, CT 06520, USA}

\author{Brandon Allen}
\affiliation{Department of Chemistry, Yale University, New Haven, CT 06520, USA}
\author{Delmar G. A. Cabral}
\affiliation{Department of Chemistry, Yale University, New Haven, CT 06520, USA}
\author{Pablo E. Videla}
\affiliation{Department of Chemistry, Yale University, New Haven, CT 06520, USA}
\author{Pouya Khazaei}
\affiliation{Department of Chemistry, University of Michigan, Ann Arbor, Michigan 48109, USA}
\author{Eitan Geva}
\affiliation{Department of Chemistry, University of Michigan, Ann Arbor, Michigan 48109, USA}
\author{Victor S. Batista}
\affiliation{Department of Chemistry, Yale University, New Haven, CT 06520, USA}

\author{Michel H. Devoret}
\altaffiliation[Present addresses: ]{Google Quantum AI, Santa Barbara, CA, USA, and Department of Physics, University of California, Santa Barbara, CA, USA}
\email{devoret@ucsb.edu}
\affiliation{Department of Applied Physics and Physics, Yale University, New Haven, CT 06520, USA}

\date{\today}
\begin{abstract}
Dissipative tunneling remains a cornerstone effect in quantum mechanics. In chemistry, it plays a crucial role in governing the rates of chemical reactions, often modeled as the motion along the reaction coordinate from one potential well to another. The relative positions of energy levels in these wells strongly influence the reaction dynamics. Chemical research will benefit from a fully adjustable, asymmetric double-well equipped with precise measurement capabilities of the tunneling rates. In this paper, we show a quantum simulator system that consists of a continuously driven Kerr parametric oscillator with a third order non-linearity that can be operated in the quantum regime to create a fully tunable asymmetric double-well. Our experiment leverages a low-noise, all-microwave control system with a high-efficiency readout, based on a tunnel Josephson junction circuit, of the which-well information. We explore the reaction rates across the landscape of tunneling resonances in parameter space. We uncover two new and counter-intuitive effects: (i) a weak asymmetry can significantly decrease the activation rates, even though the well in which the system is initialized is made shallower, and (ii) the width of the tunneling resonances alternates between narrow and broad lines as a function of the well depth and asymmetry. We predict by numerical simulations that both effects will also manifest themselves in ordinary chemical double-well systems in the quantum regime. Our work is a first step for the development of analog molecule simulators of proton transfer reactions based on quantum parametric processes.
\end{abstract}
\maketitle

\section{Introduction}\label{sec:intro}
Engineering Hamiltonians to produce a desired potential landscape is a crucial task in quantum computing \cite{wang_efficient_2020, choi_robust_2020, sun_toward_2024, lyu_mapping_2023, guo_engineering_2024, schlawin_continuously_2021}. Among these landscapes, double-wells hold particular importance, serving as models for diverse systems like two-level defects \cite{anderson_anomalous_1972, phillips_tunneling_1972}, nuclear structures \cite{merzbacher_early_2002}, and chemical reactions \cite{hund_zur_1927, schlawin_continuously_2021}. However, tuning parameters experimentally within these systems, like the barrier height, often proves to be challenging \cite{schlawin_continuously_2021, bellonzi_feasibility_2024}. Additionally, classical computational models can struggle with accuracy, e.g. failing to reach chemical accuracy \cite{simm_systematic_2016}. Consequently, developing a low-noise system capable of generating tunable double-well potentials is highly desirable for the simulation of quantum chemistry problems.

In this manuscript, we report the results of the activation dynamics of a continuously tunable asymmetric double-well parametric oscillator suitable for the simulation of chemical activation. During our exploration, we found two unexpected effects. First, we find that the asymmetric double-well can experience a significantly longer activation time (well-switching time) from one well to the other than the symmetric one, even when the system is initialized in the shallower well. This is counterintuitive because one would think that by reducing the barrier height, the activation time should decrease \cite{hanggi_reaction-rate_1990,marthaler_switching_2006,dykman_quantum_2011}. We find this is not always the case in our system due to a subtle quantum effect described below. This suggests an unexpected technique to stabilize bosonic quantum states.
The second unexpected effect is that the activation exhibits pronounced quantum resonances whose width alternates between narrow and broad with both the depth and the asymmetry of the wells. This is a manifestation of the width of the Hamiltonian anti-crossing of the energy levels close to the top of the barrier of the double-well energy surface. The location and width of these resonances are well explained by a Hamiltonian model  within the rotating wave approximation (RWA) and by a semiclassical model.

Based on the experimental and theoretical observation of these effects in the Kerr parametric oscillator (KPO), we investigate whether they can be generalized to other double-well systems and assess the capacity of our system to deal with quantum simulation applications. A particularly important class of double-well problems are found in chemistry, for example for modeling electron-transfer reactions \cite{schlawin_continuously_2021} and proton tunneling \cite{cabral_roadmap_2024}. We predict the effects should be generically observable in quantum dissipative double-wells, in particular also in double-wells that are used to model transfer reactions
like between the Guanine-Cytosine DNA base pairs. Finally, we point out that the results reported in this paper lead to the proposal that our implementation can be tuned, within realistic parameters, to quantum simulate precisely these \cite{cabral_roadmap_2024}.

\section{Setup and Model system}

\begin{figure}[t!]
    \centering 
    \includegraphics[width = \columnwidth]{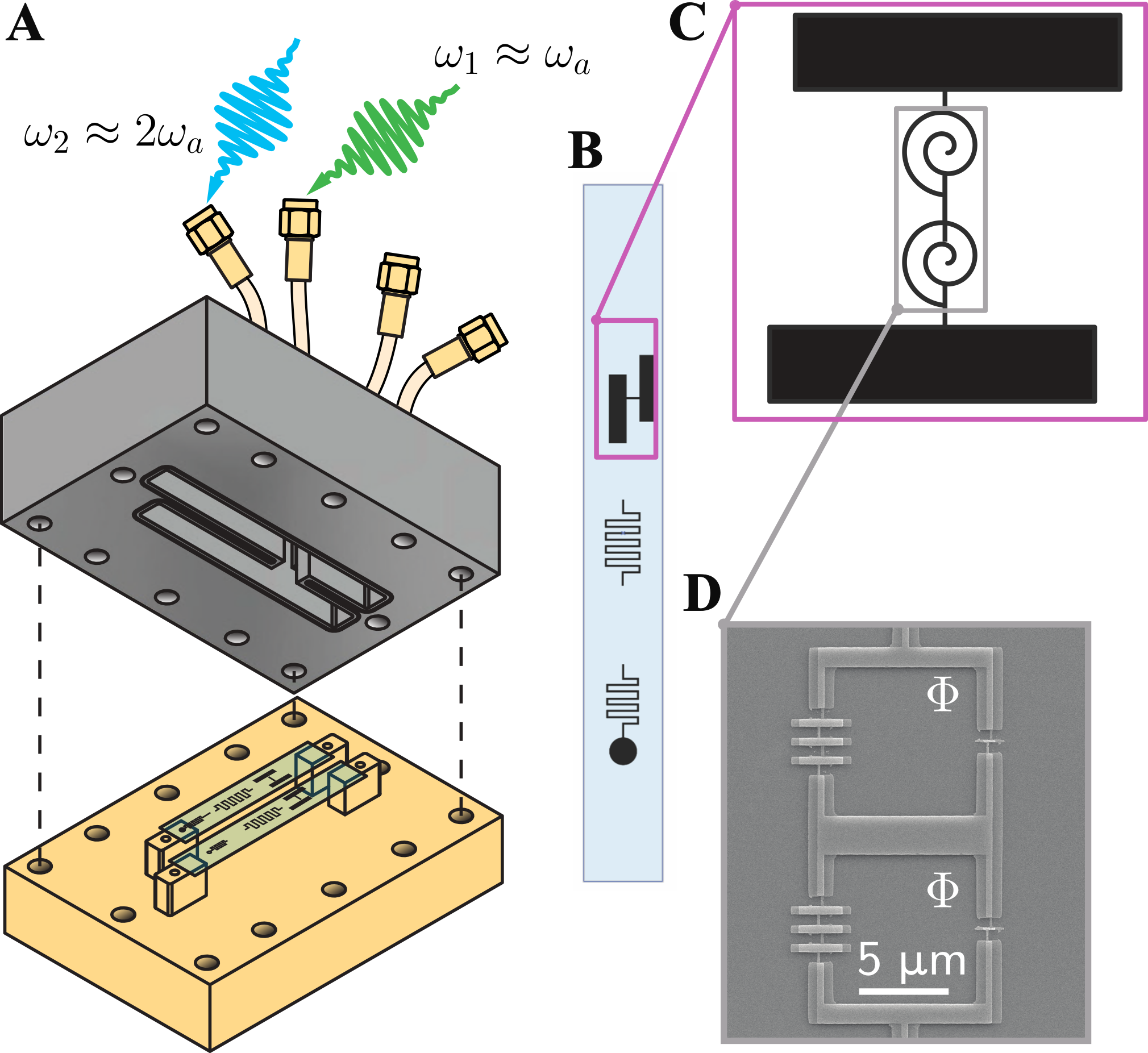}
    \caption{ \textbf{Experimental setup}. \textbf{A)} Rendering of the half-aluminum, half-copper sample package containing two sapphire chips magnified in \textbf{B)}. Each chip consists of a SNAIL-transmon, a readout resonator, and a Purcell filter. Only one chip is used in this work. The normalized resonance frequency of the SNAIL-transmon is noted $\omega_a$. Applying a strong microwave drive at $\omega_2 \approx 2\omega_a$ morphs the SNAIL-transmon Hamiltonian into the parametric oscillator Hamiltonian. \textbf{C)} Schematic of the SNAIL-transmon: a two-SNAIL array serves as the nonlinear element. The capacitor pads are shifted with respect to the axis of the array to couple it to the readout resonator. \textbf{D)} Scanning electron micrograph of the two-SNAIL array. The SNAIL loops are biased with an external magnetic flux $\Phi/\Phi_0=0.31$, where $\Phi_0$ is the magnetic flux quantum. Figure adapted from \cite{frattini_observation_2024}.}
    \label{fig:setup}
\end{figure}

Our setup was first described in \cite{grimm_stabilization_2020}, with the current implementation first introduced in \cite{frattini_observation_2024} and summarized here for the sake of completeness. The sample consists of two chips with superconducting circuits, shown in Fig.~\ref{fig:setup} \textbf{A}, that are addressable by microwave drives via charge-coupling. In this experiment, we only make use of one of the two chips, but we note that applications based on multiple devices are within the range of present day technology (see \cref{sec:chemistry}). The relevant chip contains an array of two superconducting nonlinear asymmetric inductive elements (SNAILs) \cite{frattini_3-wave_2017, frattini_three-wave_2021} shunted by a large capacitor \cite{grimm_stabilization_2020}, as depicted in Fig. \ref{fig:setup} \textbf{B}-\textbf{D}. We include a wiring diagram of the setup in the Supplementary Material (SM).

The Hamiltonian of our SNAIL transmon with charge drives can be approximated as \cite{grimm_stabilization_2020, frattini_three-wave_2021}

\begin{equation}\label{eq:H_full}
    \begin{aligned}
        \frac{\hat H(t)}{\hbar} &= \omega_o \hat{a}^{\dagger}\hat{a} + \frac{g_3}{3}(\hat{a}+\hat{a}^{\dagger})^3 + \frac{g_4}{4}(\hat{a}+\hat{a}^{\dagger})^4 \\ 
        - i\Omega_1 &\sin(\omega_1 t + \phi)(\hat{a}-\hat{a}^{\dagger}) - i\Omega_2 \sin(\omega_2 t)(\hat{a}-\hat{a}^{\dagger}),
    \end{aligned}
\end{equation}
where $\omega_o$ is the bare resonance frequency of the SNAIL transmon, $g_3,$ $g_4$ are the third- and fourth-order non-linearities of the circuit and $\hat{a}$ is the bosonic annihilation operator. This Hamiltonian is the so-called (asymmetric) parametric oscillator Hamiltonian when $\omega_2 \approx 2\omega_o$ and $\omega_1 = \omega_2/2$ \cite{ryvkine_resonant_2006,bones_resonant-force_2024}. Here, $\Omega_1$ is the amplitude and $\omega_1$ the frequency of the one-photon drive that will henceforth be referred to as the linear, or additive, drive, while $\Omega_2$ and $\omega_2$ are the amplitude and frequency of what we refer to as the squeezing, or two-photon, parametric drive. The phase $\phi$ is the relative phase between the two drives. 

By applying displaced frame transformations, transforming into the rotating frame at $\omega_2/2$ and keeping some terms beyond the RWA \cite{frattini_observation_2024,garcia-mata_effective_2024}, we arrive at the effective Hamiltonian describing the asymmetric parametric oscillator 
\begin{equation} \label{eq:H_eff}
    \frac{\hat H_{\text{eff}}}{\hbar} = \Delta \hat{a}^{\dagger}\hat{a} -K\hat{a}^{\dagger2}\hat{a}^2  + \epsilon_2\hat{a}^2  + \epsilon_1 e^{i\phi}\hat{a} + \textrm{H.c.}
\end{equation}
where  $\Delta = \omega_2/2 - \omega_a$ is the detuning and $\omega_a\approx  \omega_o$ the renormalized SNAIL transmon resonance frequency, $K = -\frac{3g_4}{2} + \frac{10g_3^2}{3\omega_a}$ is the leading order Kerr non-linearity. The drive coefficients are given by  $\epsilon_1 = \frac{\Omega_1}{2}$ and $\epsilon_2 = g_3\frac{4\Omega_2}{3\omega_a}$. 
\begin{figure}[t!]
    \centering 
    \includegraphics[width = \columnwidth]{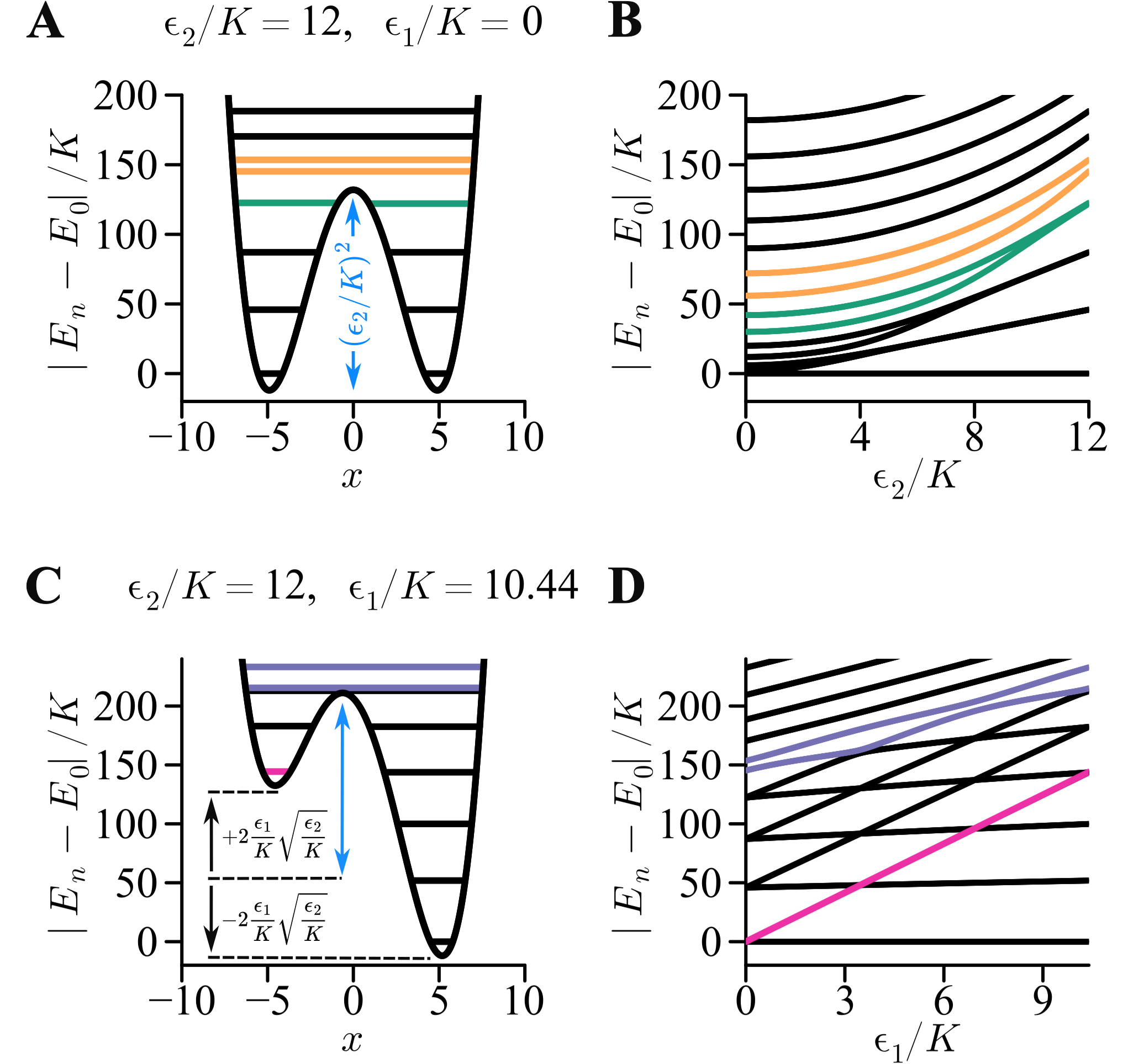} \label{fig: wells plus spectrums}
    \caption{\textbf{Symmetric and asymmetric double-well spectrum}. \textbf{A)} Effective double well potential energy $V(x)$, represented here with the quantum energy levels. \textbf{B)} Transition spectrum $|E_n-E_0|/K$ of the parametric oscillator Hamiltonian as a function of $\epsilon_2$ controlling the barrier height \cite{puri_engineering_2017}. The levels highlighted in green become degenerate at $\epsilon_2/K\approx12$, while the next pair of levels (highlighted in orange) is not yet degenerate at this value of $\epsilon_2/K$. \textbf{C)} Same as \textbf{A} but with asymmetry. \textbf{D)} The transition spectrum as a function of $\epsilon_1$ controlling the asymmetry. The ground level in the shallow well is highlighted in pink. The pair of levels at the barrier top is highlighted in violet. Note the oscillation of the energy separation between these two levels as a function of $\epsilon_1/K$.}
    \label{fig:wells}
\end{figure}
The relation to a double-well becomes apparent in the classical limit by defining 
\begin{equation}\label{eq:V(x)}
\frac{V(x)}{\hbar} = \frac{H_{\text{eff}}}{\hbar}\Big|_{ p=0 } =  k_4 x^4 - k_2 x^2 + k_1 x,
\end{equation}
where $\hat a \mapsto \frac{1}{\sqrt{2}}(x+ip)$ together with  $k_1 = \sqrt{2} \epsilon_1 \cos\phi $, $k_2 = -(\epsilon_2+\Delta/2)$, and $k_4 = -K/4$. Two instances of $V(x)$ are shown in Fig. \ref{fig:wells} \textbf{A} and \textbf{C}, while in \cref{fig:wells} \textbf{B} and \textbf{D}, the associated energy spectra of \cref{eq:H_eff} are shown as a function of the control parameters $\epsilon_1$ and $\epsilon_2$ for $\phi = 0$. From \cref{eq:H_eff} or \cref{eq:V(x)}, $\epsilon_1$ controls the asymmetry of the wells and $\epsilon_2$ controls their depth \cite{ryvkine_resonant_2006}. Note, however, that the parametric oscillator Hamiltonian cannot be written as a sum of kinetic [$T(p)$] and potential [$V(x)$] energy since cross-terms like $x^2p^2$ are present. These terms can lead to interesting consequences \cite{marthaler_switching_2006,marthaler_quantum_2007,venkatraman_driven_2023,iyama_observation_2024}. However, the perturbation of the energy levels due to these cross-terms can be decreased by reducing the zero-point spread of the oscillator. We show a rigorous derivation of the perturbative nature of these terms, as well as their impact on the simulation of chemistry, in \cite{cabral_roadmap_2024}. The perturbations due to cross-terms do not play a critical role in the new effects described in this paper, as shown by the fact that a prototypical chemical double-well Hamiltonian without such cross-terms displays the same effects as the KPO (see \cref{sec:chemistry}).

For $\epsilon_1 = 0$, we recover the conventional (symmetric) parametric oscillator Hamiltonian that creates a double-well along the position axis \cite{puri_engineering_2017,grimm_stabilization_2020, frattini_observation_2024, venkatraman_driven_2023,iyama_observation_2024, hajr_high-coherence_2024}. If $\epsilon_1 \neq 0$, the phase $\phi$ becomes relevant. For $\phi = 90\degree$, the linear drive adds a term proportional to the `momentum', $p$, thus not breaking the symmetry between the wells. For $\phi=0$, this drive adds a term proportional to the `position', $x$, thus lifting the degeneracy between the two wells \cite{ryvkine_resonant_2006,Dykman_fluctuating_2012} (see Fig. \ref{fig:wells} \textbf{C} and \textbf{D}). In the main text, we focus on the case $\phi=0$ and leave the experimental study of the effect of the phase-variation for the SM.

To model the activation rate, we use an ordinary Lindbladian model containing only single photon gain and single photon loss with phenomenological rates and temperature \cite{puri_engineering_2017,gautier_combined_2022,putterman_stabilizing_2022} that we directly fit to our data.

\section{Experiment and analysis} \label{Exper}

\begin{figure}[t!]
    \centering    \includegraphics[width=\columnwidth]{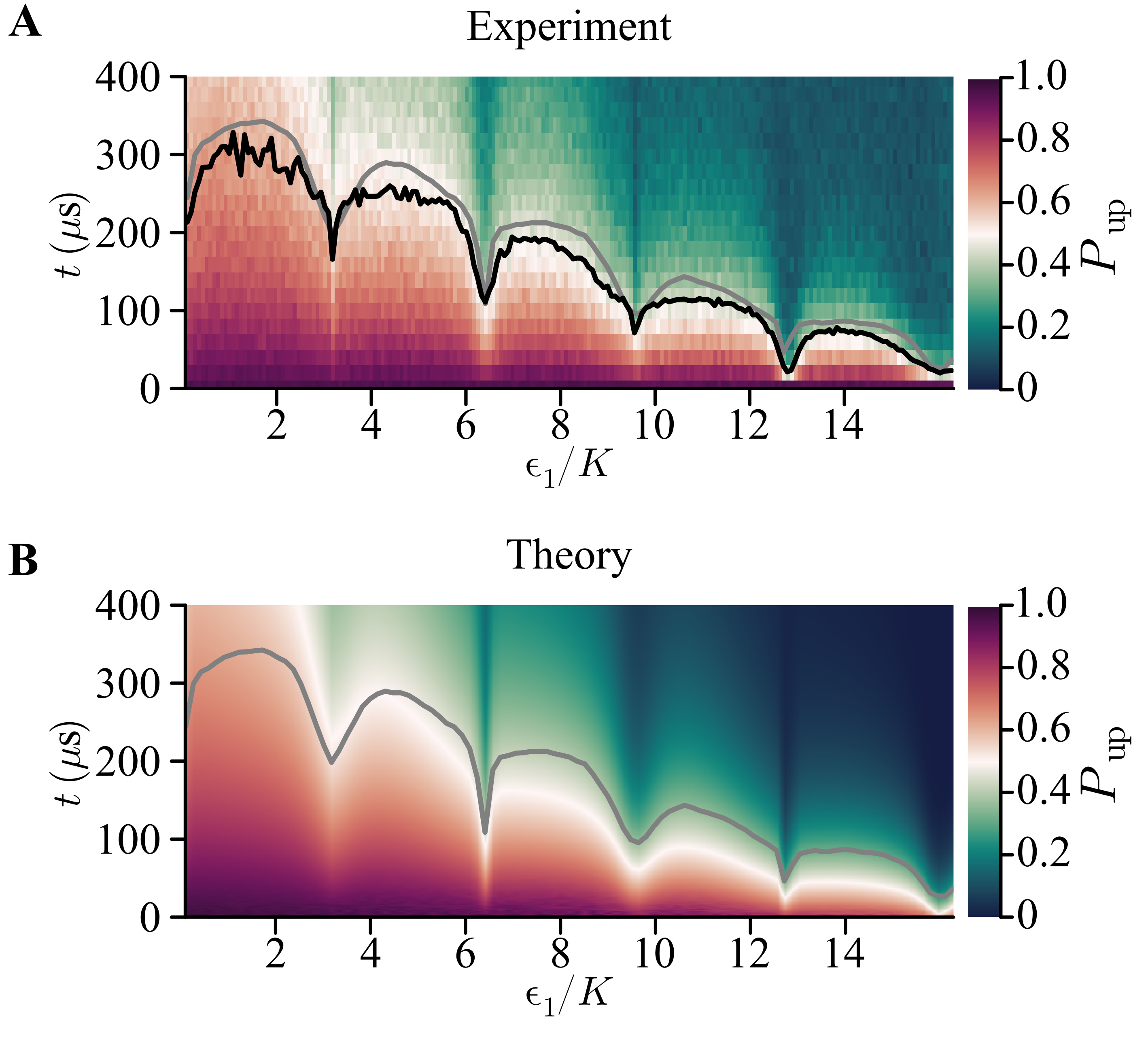}
    \caption{ \textbf{Population of the shallow well as a function of time and of the asymmetry, theory vs. experiment.} Data is shown for $\epsilon_2/K$ = 10.2. \textbf{A-B)} Probability of being in the shallow well as a function of time and $\epsilon_1/K$. The curve in black represents the decay time $\tau$ of the exponential population decay. The gray curve is the theoretical lifetime fitted from the simulation to 220~mK in \textbf{B}, overlaid for comparison.
    Resonances with characteristic widths are apparent. Notice that these widths alternate between narrow and broad. Also note that the maximum of the first lobe occurs at finite asymmetry and therefore the lifetime is larger at finite asymmetry than at zero asymmetry, which is unexpected. Data is collected and fitted from zero to 900~\textmu s but the portion from 400~\textmu s to 900~\textmu s is not shown. The theory plot in \textbf{B)} is obtained from a Lindblad model with single photon gain and loss (see main text for details). This simulation reproduces the key features observed in the experiment, including the locations of the resonances, the alternating pattern of narrow and broad widths, and the maximal lifetime occurring at small finite asymmetry.
    }
    \label{fig:exponentials}
\end{figure}

\begin{figure}[t!]
    \centering
\includegraphics[width=\columnwidth]{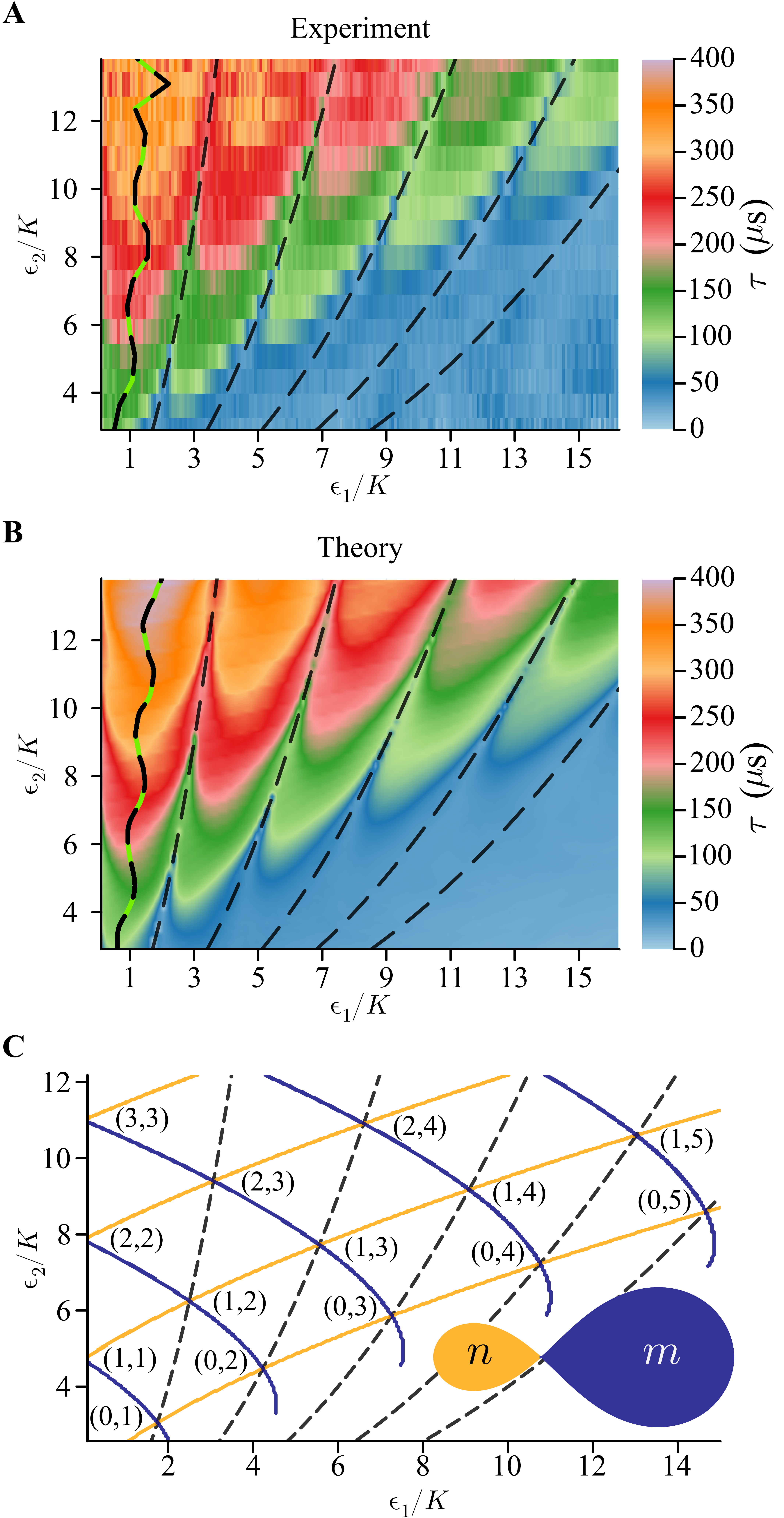}
    \caption{\textbf{Analysis of the dissipative tunneling resonances according to different models and the experiment.}
    \textbf{A)} Measurement of dissipative tunneling time $\tau$ as a function of $\epsilon_2/K$ (controlling the barrier height) and $\epsilon_1/K$ (controlling the asymmetry) in linear color scale.
    \textbf{B)} Theory fit from a Lindbladian model including single photon loss and a temperature that depends on $\epsilon_2$, growing quadratically from 170 mK to 310 mK (see SM). The wavy green-black lines show that the maxima of $\tau$ `avoid' the resonances marked by the triple intersections.
    \textbf{C)} Hamiltonian fit for the resonance conditions (parabolic dashed lines) and EBK's orbit quantization condition for $n$ and $m$ allowed quantum numbers in the small (orange) and large (blue) figure-8 lobes. The triple intersections are labeled by $(n,m)$ quantum numbers.
    }  
    \label{fig:mamba}
\end{figure}

\begin{figure}[t!]
    \centering
\includegraphics[width=\columnwidth]{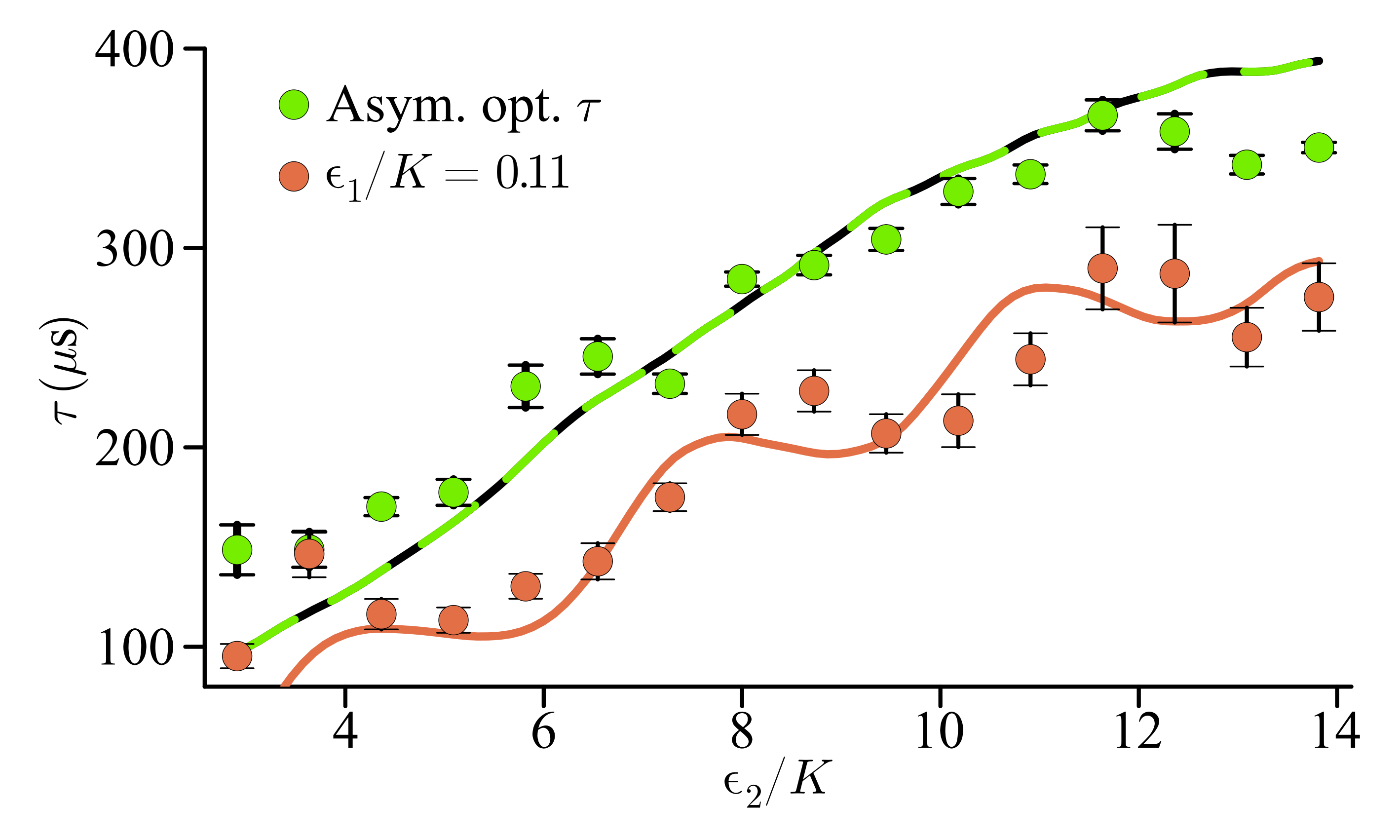} 
    \caption{\textbf{Experimental (points) and theoretical (solid lines) activation time $\tau$ for (nearly) symmetric KPO and at optimal asymmetry.} Depicted as a function of barrier height $\epsilon_2/K$. We observe that a finite asymmetry can increase the activation time. Here the asymmetric case (green) is consistently above the data for the symmetric case (orange). The asymmetric optimum corresponds to the green-black lines in \cref{fig:mamba} \textbf{A}, \textbf{B}.}
    \label{fig:black_mamba_vs_staircase}
\end{figure}

To ready the setup for our experiments, we bias our SNAIL loops with an external magnetic field sourced by a solenoid lying below the copper part of the enclosure (orange block in \cref{fig:setup} \textbf{A}). This flux sets the frequency $\omega_a$ and the Kerr nonlinearity $K$. For the theory plots in \cref{fig:mamba} and \cref{fig:black_mamba_vs_staircase}, we fit $\Delta/2\pi = 558$ kHz, and $K/2\pi = 465$ kHz to the data in \cref{fig:exponentials}. From the flux dependence of both $\omega_a$ and $K$ we can fit a model for the SNAIL \cite{frattini_three-wave_2021} to extract $g_3/3 = 2\pi \times (-5.6)$~MHz and $g_4/4 = 2\pi \times (-74)$~kHz. The values of $\Omega_1$ and $\Omega_2$ are directly proportional to the microwave amplitude we apply to our sample and, therefore, we have precise control over $\epsilon_{1}$ and $\epsilon_{2}$. Details on this calibrations are provided in the SM.

To measure the activation rate of our system, the states localized at the bottom of the wells need to be prepared and monitored as a function of time. By turning on the squeezing drive, and waiting five times the single-photon lifetime, we prepare the oscillator in its steady state. We observe the bifurcation of our SNAIL oscillator by homodyning the emitted radiation activated by a tone parametrically coupling the parametric oscillator with the on-chip readout resonator, itself coupled to our quantum-limited amplifier detection line \cite{grimm_stabilization_2020}. The homodyne signal clearly shows the typical pair of stable oscillations out-of-phase by $180^{\circ}$. Importantly, the photons emitted by the oscillator during readout are continuously replenished by the squeezing drive: the driven oscillator in presence of dissipation remains in one of its two quasi-steady states \cite{cortinas_arbitrary_2025}. Acquiring a single data point for the which-well information takes 4~\textmu s, which is typically much shorter than the activation time across the double-well parametric oscillator barrier, which is of order 100~\textmu s. We postselect the instances when the system is initially found in the shallower well. By performing a second measurement after a variable waiting time (see SM), one can detect well-switching. By repeating these measurements, one can determine the probability per unit time of witnessing an activation event and therefore obtain the activation rate for a given set of Hamiltonian parameters $\epsilon_1/K$ and $\epsilon_2/K$, which we fit using the Lindbladian model from this data.

In \cref{fig:exponentials} \textbf{A} we show measurements of the population dynamics of the wells for different values of the asymmetry, controlled by $\epsilon_1/K$. For these measurements the squeezing amplitude is set to $\epsilon_2/K =10.2$. The probability, as a function of time, of being in the initial well is well approximated by an exponential decay (see SM). This timescale $\tau$ of this exponential is a direct measurement of the activation rate ($1/\tau$) and we plot it as a black line on top of the data. In \cref{fig:exponentials} \textbf{B} we show a theory plot from a Lindbladian model fitted to the data [see SM \cref{eq:lindabladian}]. It includes single-photon loss at rate $\kappa/K = 0.011$ and gain associated with a temperature of 220~mK. The theory fit has an associated activation time plotted as a gray line in both \cref{fig:exponentials} \textbf{A} and \textbf{B}.

We observe unexpected resonances for certain values of $\epsilon_1/K$, where the activation rate is markedly increased. These are resonances between levels localized in different wells. The resonances for levels deep within the wells behave effectively as level crossings since the coupling is exponentially small due to the suppression of tunneling under the barrier. Therefore, these activation events occur through a two-step process: population is first transferred from the bottom of the well to eigenstates at the barrier-top (i.e. those whose energies lie near the local maximum of the double-well potential) by thermal and quantum heating \cite{dykman_quantum_2011,marthaler_switching_2006}, and then tunnels to the other well \cite{frattini_observation_2024}. An extensive theoretical analysis of the different heating mechanisms and the importance of the barrier-top eigenstates in KPOs is provided in \cite{su_unraveling_2024}. Briefly summarized, thermal heating drives population into barrier-top superposition states by absorbing real photons, while quantum heating arises because photon loss in the Fock basis can appear as heating when viewed in the eigenbasis of the effective Hamiltonian. In our parameter regime, thermal heating dominates \cite{su_unraveling_2024}.
That the crossing is of over-the-barrier type is also seen from \cref{fig:exponentials,fig:mamba}  by noting that the alternating width of the different resonance in \cref{fig:wells} \textbf{D} correspond to the strength of the anticrossings of the spectrum at the barrier top (purple levels in \cref{fig:wells} \textbf{C}). This interpretation of the data allows us to predict the location in parameter space where resonant tunneling takes place. This happens when the uncoupled levels in the right and left well align. By realizing that the energy spacing of the levels in the wells can be estimated by $S \approx 4\epsilon_2$ \cite{puri_engineering_2017} and that the asymmetry, defined as the energy difference between the lower-lying state of each well, can be estimated from \cref{eq:V(x)} as $A \approx 4\epsilon_1\sqrt{\epsilon_2/K}$ (see also \cite{bones_resonant-force_2024}) we write the resonance condition as ($A=nS$)
\begin{equation} \label{eq:parabola}
    \frac{\epsilon_2}{K} \approx \left( \frac{\epsilon_1}{nK} \right)^2,
\end{equation}
where $n$ numbers the different resonances. Whenever the levels inside the wells align, the levels close to the barrier top also align, which facilitates the passage from one well to the other. This simple formula dictates the location of the resonances (see dashed lines in \cref{fig:mamba}).

We complement this analysis with a semiclassical action quantization taken here as a proxy for the quantum levels. The number of allowed quantum orbits is given by the number of action quanta enclosed by the asymmetric ``figure eight'' lemniscate delineating the phase space separatrix in between the wells. The orange and blue curves in \cref{fig:mamba} \textbf{C} show the pairs $\epsilon_1/K$ and $\epsilon_2/K$ where the Einstein–Brillouin–Keller (EBK) \cite{noauthor_einsteins_nodate} action quantization condition is met. That is, 
\begin{equation} \label{eq: semi-classical quantisation}
\frac{1}{2\pi} \oint \ p\,dx=  \hbar \left(\tilde{n} + \frac{1}{2}\right)
\end{equation}
where $\tilde{n} = n,m$ are the quantum numbers of the shallow and deep wells respectively. The triple intersection points of the parabolas from \cref{eq:parabola} with the equi-action curves meeting the EBK quantization condition mark the point in parameter space where a new level enters the wells in the tunneling resonances condition. We mark the triple intersections in \cref{fig:mamba} \textbf{C} by $(n,m)$, the quantum numbers of each well. At these points, where each well contains exactly $n$ and $m$ semiclassical orbits, we expect the resonance to broaden due to a new orbit contributing to the activation rate. This is seen as sharpening structures in \cref{fig:mamba} \textbf{A} and \textbf{B} (see also \cref{fig:exponentials} for the broadened resonances).

In \cref{fig:mamba} \textbf{A} we show the measured activation time for a scan of both $\epsilon_1/K$ and $\epsilon_2/K$. The resonances are shown to have widths that change along the scan. They are found to reflect the width of the tunneling anti-crossings at the top of the barrier. This can be seen by following the purple energy curves and their tunnel splitting in \cref{fig:wells} \textbf{D}. The semiclassical Hamiltonian theory predicts, quantitatively, the resonance condition and, qualitatively, their widths (see SM for a full quantum treatment). To obtain absolute activation times, we employ a Lindblad model rather than the semiclassical EBK theory.

In \cref{fig:mamba} \textbf{B}, we show the  Lindbladian model fit to the experiment. The fit requires a temperature dependence with $\epsilon_2$ which is well approximated by a quadratic function increasing from $\sim 170$~mK at $\epsilon_2\approx3$ to $\sim310$~mK at $\epsilon_2/K\approx14$ (see SM for the temperature extraction). This is consistent with Johnson-Nyquist thermal noise produced by the heating of the attenuators with increasing squeezing amplitude. However, as we expand in the SM, we note that this is merely phenomenological and more involved physics are possibly involved. Other sources for the effective temperature can be unwanted mixing Josephson terms \cite{venkatraman_nonlinear_2024} or other modes in the system \cite{benhayoune-khadraoui_how_2025}. A way to mitigate this temperature exists, however, and is being developed and has been already implemented elsewhere by us \cite{ding_quantum_2025} and our colleagues \cite{adinolfi_enhancing_2025}.

We learned from this data that an asymmetric system can have a longer activation lifetime than the symmetric system, even if one of the wells is markedly shallower (see \cref{fig:exponentials} at $\epsilon_1/ K\approx1$). The green-black dashed line in \cref{fig:mamba} \textbf{A} shows the experimentally determined maxima of activation time $\tau$ as a function of the control parameters. The theoretical maxima are shown in \cref{fig:mamba} \textbf{B} by the line with the same colors and agree with those we find by keeping the temperature constant in the model (not shown). The location of the maxima is nontrivial, and it is relevant to control reaction rates, or analogously because it can readily be used to extend the lifetime of a Kerr-cat qubit \cite{grimm_stabilization_2020}. 

In \cref{fig:black_mamba_vs_staircase}, we show the increase of $\tau$ with the asymmetry parameter. To provide physical insight into this effect, we note that Lindblad theory predicts that $\tau$ is modulated in a step-like fashion as the quantized orbits fall under the barrier \cite{frattini_observation_2024} (see also SM). The linear drive can be exploited to break the parity symmetry of these orbits as shown in \cref{fig:wells} and therefore avoid the step-like modulation seen in the Lindblad theory for $\epsilon_1 \approx 0$. 
This observation explains the trajectory of the green-black optimal curve in \cref{fig:mamba} \textbf{A}, \textbf{B}, which avoids the new resonance conditions marked by the triple intersections. In other words, quantum tunneling produces a hybridization of the classically decoupled orbits right under the barrier. The asymmetry parameter can be adjusted to minimize that hybridization, reducing the tunneling rate via the excited states.

\section{Quantum simulation of chemistry}
\label{sec:chemistry}
\begin{figure}[t!]
    \centering
\includegraphics[width=\columnwidth]{figs/chemical_scars_plus_b_mamba_with_tau.png}
    \caption{\textbf{Lindbladian simulation of an ordinary double-well system}. \textbf{A)} Dynamics simulated as a function of well-asymmetry $k_1/k_4$ and well-depth $k_2/k_4$. The color code marks the activation time (compare to Fig. \ref{fig:mamba} \textbf{B}). The green-black line shows the maximum activation time and the orange line marks the symmetric case ($k_1=0$). \textbf{B)}  Comparison of the activation time for the ordinary symmetric ($k_1=0$) double-well and the (optimal) activation time along the green-black line in \textbf{A} as a function of well-depth $k_2/k_4$. \textbf{C)} Activation time as a function of asymmetry for $k_2/k_4=12.6$ (black dashed line in \textbf{A}), showing resonant tunneling (as in Fig. \ref{fig:exponentials}, see also \cite{schlawin_continuously_2021}). Note that, as in the parametric oscillator \cref{fig:exponentials}, the linewidths alternate from broad to narrow.}
    \label{fig:chemical black mamba}
\end{figure}

The experimental observations presented here raise the question of whether they are present in other double-well systems, like the type of double-well involved in modeling chemical reactions. In chemistry, while harmonic modes coupled to spins are often used to model charge transfer in a simplified framework \cite{schlawin_continuously_2021}, a true double-well description offers a more natural representation in the adiabatic regime, where the potential energy landscape plays a crucial role. The true double-well potential provides a general framework that accurately captures essential features of numerous chemical processes, such as proton transfer, isomerization, and proton-coupled electron transfer \cite{cabral_roadmap_2024}.

We run a Lindbladian simulation for the chemical system to investigate if our experimental observations also manifest here. For this, we use the chemically inspired Hamiltonian $\hat{H}/ \hbar = \hat{p}^2/2 + k_4 \hat{x}^4 - k_2 \hat{x}^2 + k_1 \hat{x}$, including single photon loss and gain with rate $\kappa/k_4 = 0.025$ and at constant temperature $n_\mathrm{th}= 0.05$. These parameters lie in the ranges accessible by our system. The Lindblad approach is valuable because it produces chemical dynamics with well-defined reaction rates, and qualitatively agrees with experimental observations in the weak coupling regime \cite{jean_application_1992, ishizaki_adequacy_2009, schlawin_continuously_2021, vu_computational_2025}. In e.g. electron-transfer chemistry, a typical choice of dissipation is Ohmic noise, which under standard approximations translates into single-photon loss and gain represented by the dissipators of the Lindblad model \cite{so_trapped-ion_2024}. 

With this framework, we answer our question affirmatively: our experimental observations are predicted to be present in other double-well systems. On the one hand, the green-black line in Fig. \ref{fig:chemical black mamba} \textbf{A} depicts the maximum lifetime as a function of well-asymmetry and well-depth, showing that - just like in our experiment (Fig. \ref{fig:mamba} \textbf{A}, \textbf{B}) - a small asymmetry increases the lifetime significantly (see also Fig. \ref{fig:black_mamba_vs_staircase} and Fig. \ref{fig:chemical black mamba} \textbf{B}). 
On the other hand, \cref{fig:chemical black mamba} \textbf{C} depicts a horizontal linecut of Fig. \ref{fig:chemical black mamba} \textbf{A} and clearly reveals the same width alternation (broad-narrow-broad) as discovered and explained in our experiment (see Fig. \ref{fig:exponentials}).
The finding of these two unexpected effects shows that our asymmetric Kerr parametric oscillator setup is already able to produce meaningful predictions for chemical quantum rate theory. 

Double-well systems have been extensively studied in various contexts \cite{so_trapped-ion_2024, sias_resonantly_2007}, including in superconducting circuits \cite{devoret_resonant_1984, yu_energy_2004, dutton_electromagnetically_2006, yu_multi-photon_2005}. However, we speculate that the unique combination of precise real-time microwave control, complete tunability over Hamiltonian parameters, experimental stability, fast repetition rates, and high-fidelity readout in our setup explains why we observed effects not previously reported.

 Based on these results, we proposed a hardware modification to our setup that implements a one-to-one single-transmon quantum simulation of tautomerization reactions in Malonaldehyde (cis-cis) and proton transfer reactions between the DNA base pairs Guanine-Cytosine \cite{cabral_roadmap_2024}. The key to this simulator is that, for \textit{realistic circuit parameters}, the Hamiltonian cross terms $\propto x^2p^2$ and the relativistic-like $\propto p^4$ term become irrelevant perturbations [these arise from the Kerr term $\hat a ^{\dagger 2}\hat a^2$, which in turn arises from the $x^4$ term as perturbations, see \cref{eq:H_full}]. Also, our system allows for a clean microwave control of effective temperature. Injecting noise through a microwave drive port is an established method in circuit QED to increase the effective temperature ($n_{th}$ in the Lindblad model) of a system \cite{slichter_measurement-induced_2012}. We emphasize that our microwave tones are generated using arbitrary waveform generators and can therefore create nearly arbitrary noise spectra. Controllable single-photon dissipation has been demonstrated for KPOs \cite{frattini_observation_2024, ding_quantum_2025}. By selecting the frequency of the dissipated photons, this mechanism can be used to engineer cooling (reduction of $n_{th}$) \cite{ding_quantum_2025}, or to increase the single-photon loss rate ($\kappa$ in the Lindblad model) \cite{frattini_observation_2024}. Beyond this, exotic forms of dissipation have been proposed \cite{venkatraman_nonlinear_2024, putterman_stabilizing_2022}. Finally, controllable interactions between two KPOs have been generated in our setup, which can create non-Markovian dynamics \cite{schaefer_aps_2024}. All of this allows for the experimental exploration of chemical dynamics across a wide range of parameter spaces \cite{friedman_quantum_1998,schlawin_continuously_2021} in this new type of single-transmon parametric quantum simulator.

We conclude these section by giving a perspective for how this approach can be generalized beyond a single SNAIL-transmon. Controllable interactions of form $\hat{a}_1^{\dagger}\hat{a}_2 + h.c.$ have been implemented for KPOs in superconducting circuits \cite{hoshi_entangling_2025}. This interaction translates to a bilinear coupling $\hat{x}_1\hat{x}_2$, opening promising avenues for applications in chemistry. The description of concrete chemical systems that can be mapped to multiple coupled KPOs is the subject of ongoing work and will be presented in a forthcoming manuscript. Furthermore, the superconducting platform itself supports strong scalability, with current systems already exceeding 1000 transmons \cite{noauthor_ibm_2023}.

\section{Conclusion}
\label{sec:conc}

We reported the measurement of the activation rate in a continuously tunable asymmetric Kerr parametric oscillator with dissipation and observed a fine structure that, to the best of our knowledge, was unknown in the literature. Our experiment shows that the activation rate displays resonances whenever a level close to the barrier top aligns with one in the other well. We derive an analytical formula that predicts the occurrence of these resonances as a function of asymmetry and well depth. Furthermore, we discover that these tunneling resonances alternate in width between narrow and broad lines as the asymmetry and well depth are changed. We trace this effect back to the alternating strengths of level anticrossings in the spectrum close to the barrier top. This shows
that the activation is of over-the-barrier type (i.e., not
via direct quantum tunneling by the low-lying states),
as predicted \cite{dykman_quantum_1988,marthaler_switching_2006,bones_resonant-force_2024}.
We are thus able to learn the level structure near the barrier top without having to prepare these excited states.

Our control over the well-asymmetry indicates that quantum parametric oscillators can implement analog quantum simulation of chemical reaction dynamics \cite{cabral_roadmap_2024}. This allows for instance the analog simulation of proton tunneling, and e.g. the study of transfer reactions between the Guanine-Cytosine DNA base pairs, appears within reach of current Kerr parametric oscillator technology \cite{cabral_roadmap_2024}. 

We also note the importance of this system for quantum computation since qubits can be encoded in the well state manifold \cite{puri_engineering_2017, grimm_stabilization_2020}. In this regard, two contributions of the present work deserve to be highlighted. The first one is the increase of the activation timescale $\tau$ by a fine control of the asymmetry. This leads to a reduction of bit-flip errors \cite{darmawan_practical_2021} with no extra hardware requirements. The second contribution is the demonstration of the operation of a highly asymmetric parametric oscillator in the quantum regime. We provide direct evidence that the static effective description is not compromised under strong linear drives, which are required for fast gates \cite{puri_engineering_2017} and new implementations of hardware efficient readout schemes \cite{suzuki_quantum_2023}. 

After writing this manuscript, we were made aware of a similar experiment at Rice University using trapped-ions to create an asymmetric double-well to simulate electron transfer 
\cite{so_trapped-ion_2024}.

\section*{Author contributions}

ACCdA, RGC, and MS analyzed the data and modeled the experiment. ACCdA, RGC, MS, and MHD wrote the manuscript. PEV supported the data analysis. All authors contributed to the manuscript. RGC and MS conceptualized the experiment and collected data. NEF fabricated the device. BA, DGAC, PK, EG, and VSB validated the relevance and utility to chemistry and provided theory support. MS coordinated the interdisciplinary collaboration and led the revision process. MHD and RGC supervised the research. 

\section*{Acknowledgements}
We acknowledge useful discussions with Q. Su, A. Ding, B. Brock and M. Dykman. Remarks by V. Joshi and A. Koottandavida have improved the manuscript. This research was sponsored by the Army Research Office (ARO) under award number W911NF-23-1-0051. The views and conclusions contained in this document are those of the authors and should not be interpreted as representing the official policies, either expressed or implied, of the Army Research Office (ARO), or the U.S. Government. The U.S. Government is authorized to reproduce and distribute reprints for Government purposes notwithstanding any copyright notation herein.
VSB, EG and MHD acknowledge partial support from the National Science Foundation Center for Quantum Dynamics on Modular Quantum Devices (CQD-MQD) under Award Number 2124511.

\section*{Supplementary Material}
\subsection{Experimental setup}
\label{expSetup}
\begin{figure*}[t]
    \centering
    \includegraphics[width=\columnwidth*2]{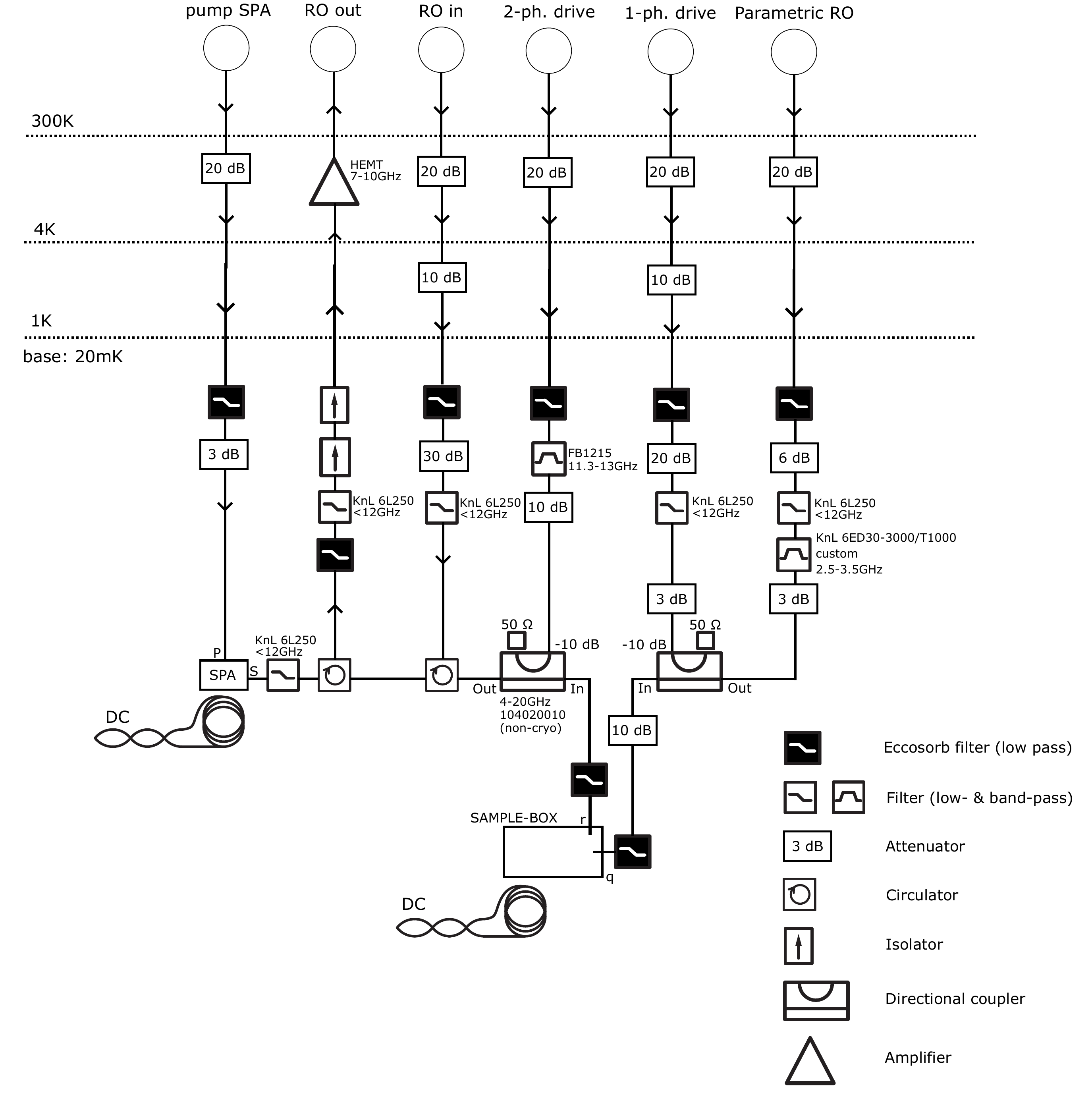}
    \caption{\textbf{Wiring diagram.} A legend for the main elements is provided at the bottom right. The approximate temperatures for different stages are shown on the left. The sample is contained in the sample box.} 
    \label{fig: fridge}
\end{figure*}

The wiring diagram is shown in \cref{fig: fridge}. The lines from left to right are the following: 1. the microwave pump for the SNAIL-Parametric-Amplifier (SPA) \cite{frattini_optimizing_2018}, 2. the readout output line, 3. the input line for the dispersive readout, 4. the two-photon drive at frequency $\omega_2 \approx 2\omega_o$, 5. the one-photon drive at frequency $\omega_1 = \omega_2/2$, 6. the input line for the parametric readout, which is the readout method utilized in this manuscript. Additionally, there are two twisted-pair cables that provide a DC signal to magnet spools that create a magnetic flux for the SPA and the SNAIL-transmon sample. Our setup is similar to the one first described in \cite{grimm_stabilization_2020}.

\subsection{Calibrations of experimental parameters}
\subsubsection{Measurement of Kerr coefficient and frequency} \label{KerrCal}
\begin{figure}[t!]
    \centering
\includegraphics[width=\columnwidth]{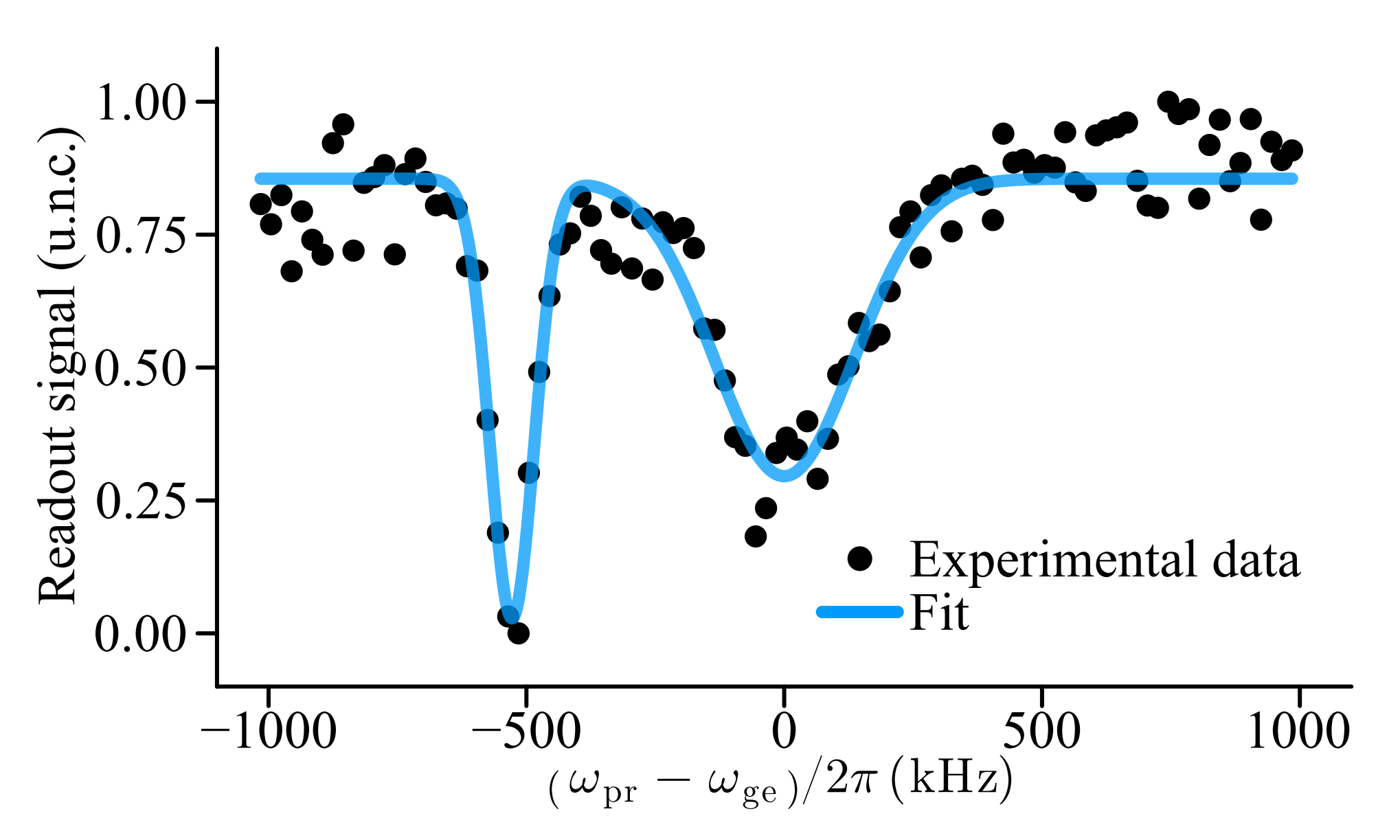}
    \caption{\textbf{Continuous-wave spectroscopy measurement showing the readout response as a function of the probe tone frequency}. From left to right, the pronounced dips in the signal show the $gf/2$ and $ge$ transitions of the SNAIL transmon. Those occur at $(\omega_{ge} - K)/2\pi$ and $\omega_{ge}/2\pi$ respectively. Fitting to this experimental data, we extract a Kerr value of $K/2\pi = (528 \pm 10)$ kHz. However, this value is not used in any of the simulations (theory fitted $465$ kHz is used instead).}
    \label{fig: Kerr fit}
\end{figure}

\begin{figure}[t!]
    \centering
\includegraphics[width=\columnwidth]{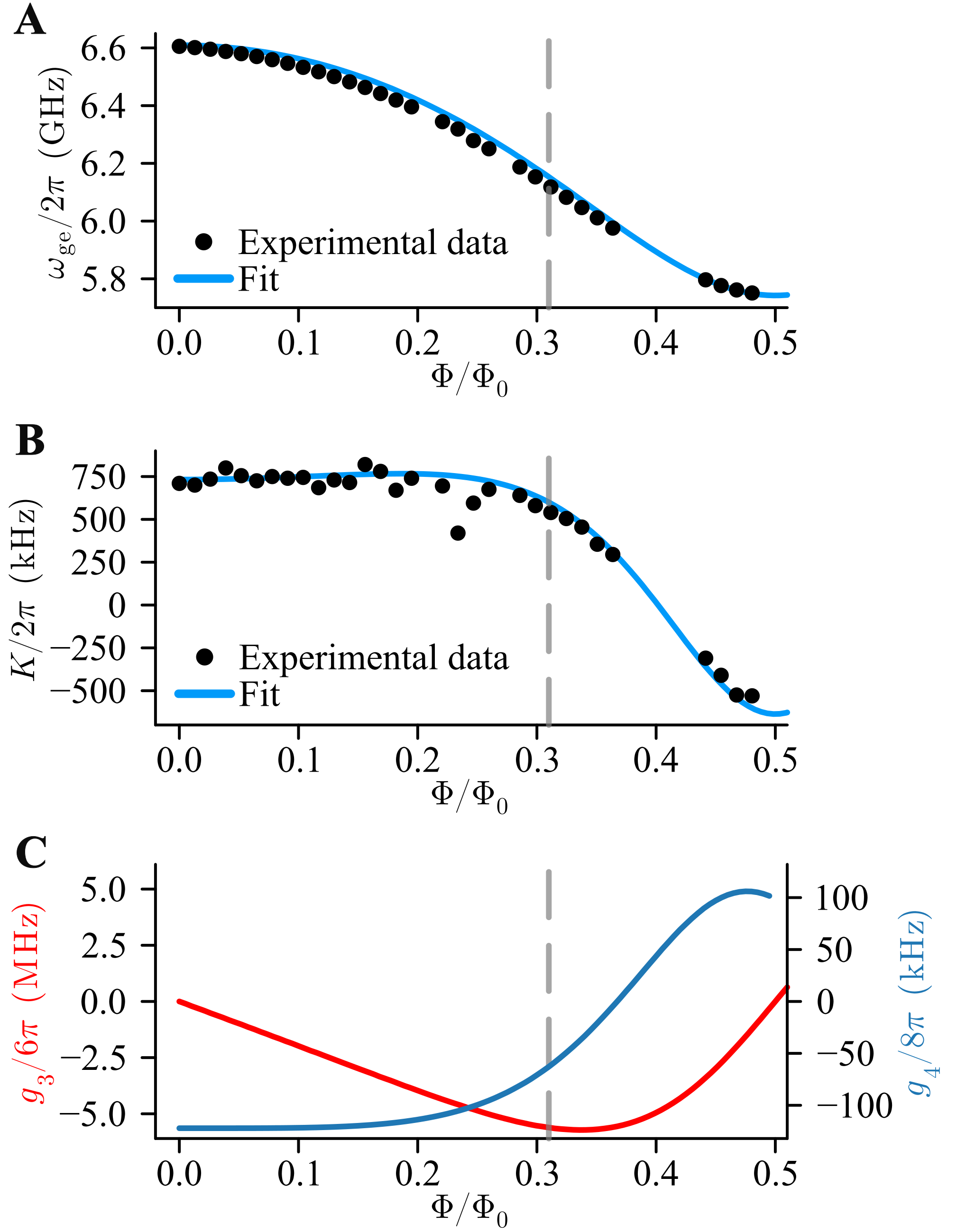}
    \caption{\textbf{Extraction of non-linearities from fit to frequency and Kerr parameter.} \textbf{A)} Spectroscopic measurement of the oscillator's frequency and \textbf{B)} its Kerr non-linearity as a function of flux as extracted from the $ge$ and $gf/2$ transitions (see \cref{fig: Kerr fit}). Solid lines are a simultaneous fit of frequency and non-linearity from the SNAIL circuit model \cite{frattini_three-wave_2021}. We remark that several arrangements of parameters fit the data satisfactorily, producing a large correlation and errors. In \textbf{C)} we show the third and fourth order non-linearities as extracted from the fit. The vertical dashed line represents the flux operating point for the experiments in the main text.}
    \label{fig: g4g3}
\end{figure}

The Kerr coefficient $K$ is extracted by spectroscopy. A saturating probe drive $\omega_{pr}$ is applied to the SNAIL transmon operated with $\epsilon_2 = 0$. Varying the probe drive frequency and measuring the response via dispersive readout, results in the data shown in \cref{fig: Kerr fit}. The spectrum shows two clear dips, which correspond, from lower frequency to higher frequency, to the two-photon $gf/2$ transition and the $ge$ transition (the SNAIL levels are labeled following the atomic physics convention in increasing energy order $g,\;e,\;f$). Note that the measured $\omega_{ge}$ is a good approximation for $\omega_a$ used in the main text (\cref{eq:H_eff}). We then extract the Kerr frequency by using the relation $\omega_{ge} - \omega_{gf/2} = K$. Fitting two Gaussian peaks to the spectrum in \cref{fig: Kerr fit}, we find $K/2\pi = (528 \pm 10)$ kHz. This measured value differs from the fitted one ($465$ kHz) by 12\% and it is not used in the simulations.
These measurements, when taken as a function of the biasing flux $\Phi$ of the superconducting loops, allow for a calibration of the SNAIL non-linear parameters $g_3, g_4$. In \cref{fig: g4g3} we show a flux scan of our sample with measured frequencies, Kerr nonlinearities, and the fit by the model presented in \cite{frattini_three-wave_2021}.

\subsubsection{Calibration of relative phase between squeezing drive and linear drive}
\begin{figure}[t!]
    \centering   
    \includegraphics[width=\columnwidth]{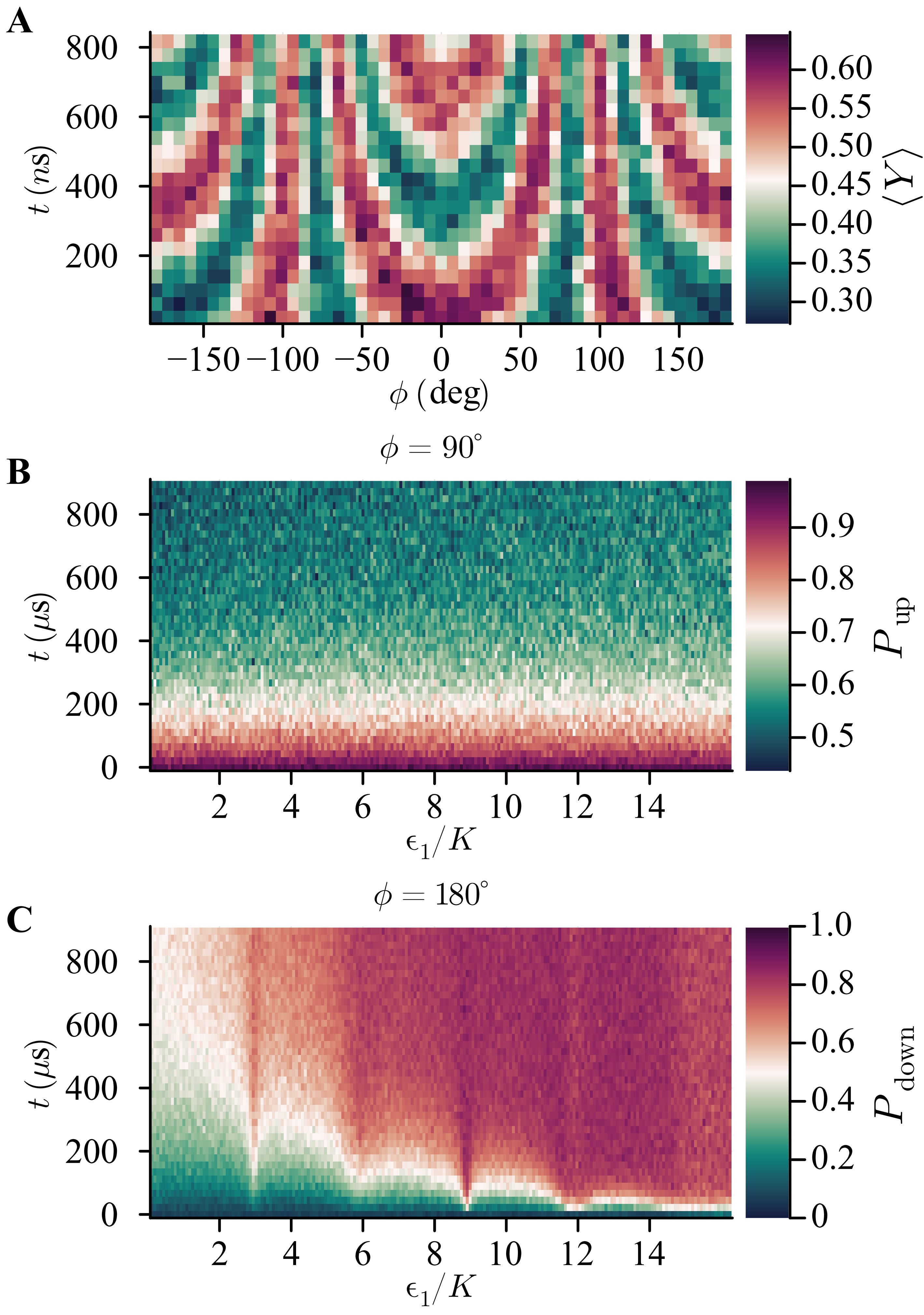}
    \caption{\textbf{Effects of the relative phase of the linear drive.} \textbf{A)} Time-resolved quantum coherent oscillation in $\hat{Y} = i\ket{\alpha}\bra{-\alpha} - i\ket{-\alpha}\bra{\alpha}$ as a function of relative phase $\phi$ between the squeezing and linear drives \cite{grimm_stabilization_2020}. This measurement shows Rabi-like oscillation between the cat states created by superposing states in different wells. The oscillation frequency is a direct measure of the asymmetry. \textbf{B)} Data taken for $\epsilon_2/K = 8.7$. Linear drive with relative phase of $\phi = 90 \degree$. The wells are slightly deformed but they remain mirror symmetric to each other and no resonances are visible. \textbf{C)} Data taken for $\epsilon_2/K = 8.7$. As in \cref{fig:exponentials} \textbf{A}, we measure the activation rate for different asymmetries, but now caused by a linear drive with a phase $\phi=180\degree$ instead of $\phi=0\degree$. The flipped phase results in an exchange of the left and right well, meaning that now the left well is the deeper one. In the experiment, we still initialize in the shallower (now right) well, but now measure the time-resolved population of the deeper (now left) well. As a result, the population of the left well is initially zero and then increases over time. We again observe trends and features, such as the resonant-tunneling also seen in \cref{fig:exponentials} \textbf{A}. This confirms that the $180\degree$ phase shift only exchanges the roles of the wells, but otherwise exhibits the same physical phenomena.}
    \label{fig:cp03phase 180}
\end{figure}

Quantum coherent Rabi-like oscillations as a function of this phase $\phi$ are shown in \cref{fig:cp03phase 180} \textbf{A} \cite{grimm_stabilization_2020}. For $\phi = 0$, this drive is position-like and lifts the degeneracy between the two wells (see Fig. \ref{fig:wells} \textbf{C} and \cref{fig: monster appendix} \textbf{A}). This energy difference induces the Rabi-like oscillation. For $\phi = 90\degree$ (\cref{fig:cp03phase 180} \textbf{A}), the linear drive is momentum-like (the Hamiltonian term is $\propto \hat p$),
thus not breaking the symmetry between the wells (see
Fig. \ref{fig: monster appendix} \textbf{D}). The well-asymmetry is determined by $\epsilon_1\cos{\phi}$, and is therefore first-order insensitive to phase drifts of a few degrees around $\phi = 0$, which is the precision of our calibration.

\subsubsection{Calibration of the parametric squeezing drive amplitude $\epsilon_2$ and linear drive amplitude $\epsilon_1$} \label{ampCal}
\begin{figure}[t!]
    \centering
    \includegraphics[width=\columnwidth]{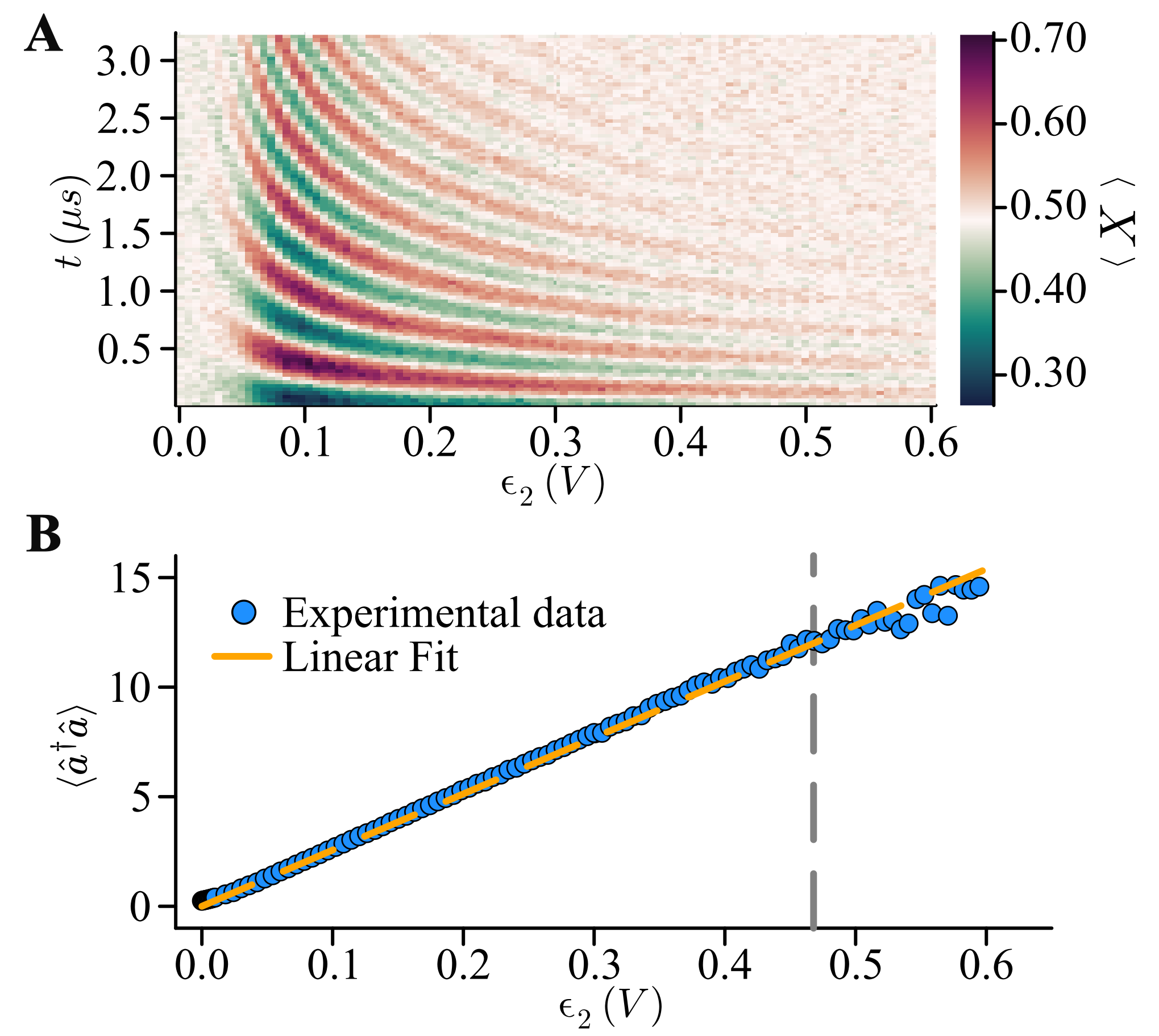}
    \caption{\textbf{Calibration of squeezing drive strengths.} \textbf{A)} Time-resolved quantum coherent Rabi-like oscillations as a function of squeezing amplitude. The squeezing amplitude is controlled as the voltage of the digital controller. \textbf{B)} Photon number $\langle \hat{a}^\dagger \hat{a} \rangle$ as function of applied voltage for the digital control of squeezing drive $\epsilon_2$. The experimental data points are obtained from \cref{fig:rabi plus nbar calib} \textbf{A} using $\langle \hat{a}^\dagger \hat{a} \rangle = \epsilon_2/K = \Omega_{\mathrm{Rabi}}^2/16\epsilon_1^2$, where $\epsilon_1$ is the asymmetry for $\phi = 0$. A linear fit allows us to convert the voltage set by the digital control of $\epsilon_2$ to the squeezing drive $\epsilon_2$ in MHz. In the main text, we only present data measured with up to $\epsilon_2/K = 13.8$ (vertical dashed gray line).
    }
    \label{fig:rabi plus nbar calib}
\end{figure}

In the experimental setup, the drive amplitudes are directly controlled by a digital to analog voltage converter. To calibrate the strength of the drive in MHz we measure time-resolved Rabi oscillations as a function of the digital control of the squeezing drive $\epsilon_2$. The experimental data is shown in Fig. \ref{fig:rabi plus nbar calib} \textbf{A}. The oscillations of the observable $\hat{X} = (\ket{\alpha}\bra{\alpha} - \ket{-\alpha}\bra{-\alpha})$
occur at a rate $\Omega_{\mathrm{cat}}(\epsilon_2) \approx \Re(4\epsilon_1\alpha^{*})$ where $\alpha = \sqrt{\epsilon_2/K}$ and the approximation is valid for $|\alpha|>1$. Using, also, that for $\epsilon_2 = 0$ the oscillation has a frequency of $2\epsilon_1$ \cite{grimm_stabilization_2020}, just like for an ordinary transmon, we obtain $\epsilon_1$ in MHz completing the calibration of the drive amplitudes. To be clear, we can rewrite this relation as $\langle \hat{a}^\dagger \hat{a} \rangle = \epsilon_2/K = \Omega_{\mathrm{Rabi}}(\epsilon_2)^2/16\epsilon_1^2$, where $\langle \hat{a}^\dagger \hat{a}\rangle$ is the average photon number of the coherent states. By extracting the Rabi rate $\Omega_{\mathrm{cat}}$ for each voltage of the digital control of $\epsilon_2$ and using the previously determined value of $\epsilon_1$, we find $\langle \hat{a}^\dagger \hat{a} \rangle$ as a function of the digital control of $\epsilon_2$. The data is shown in Fig. \ref{fig:rabi plus nbar calib} \textbf{B} and shows a linear relationship between the applied voltage for the drive and the average photon number. The slope of the linear fit (together with the previously extracted value of Kerr) determines the proportionality constant between the digital to analog converter in volts and the drive amplitude $\epsilon_2$ in MHz as required by the Hamiltonian description.

\begin{figure}[t!]
    \centering
    \includegraphics[width=\columnwidth]{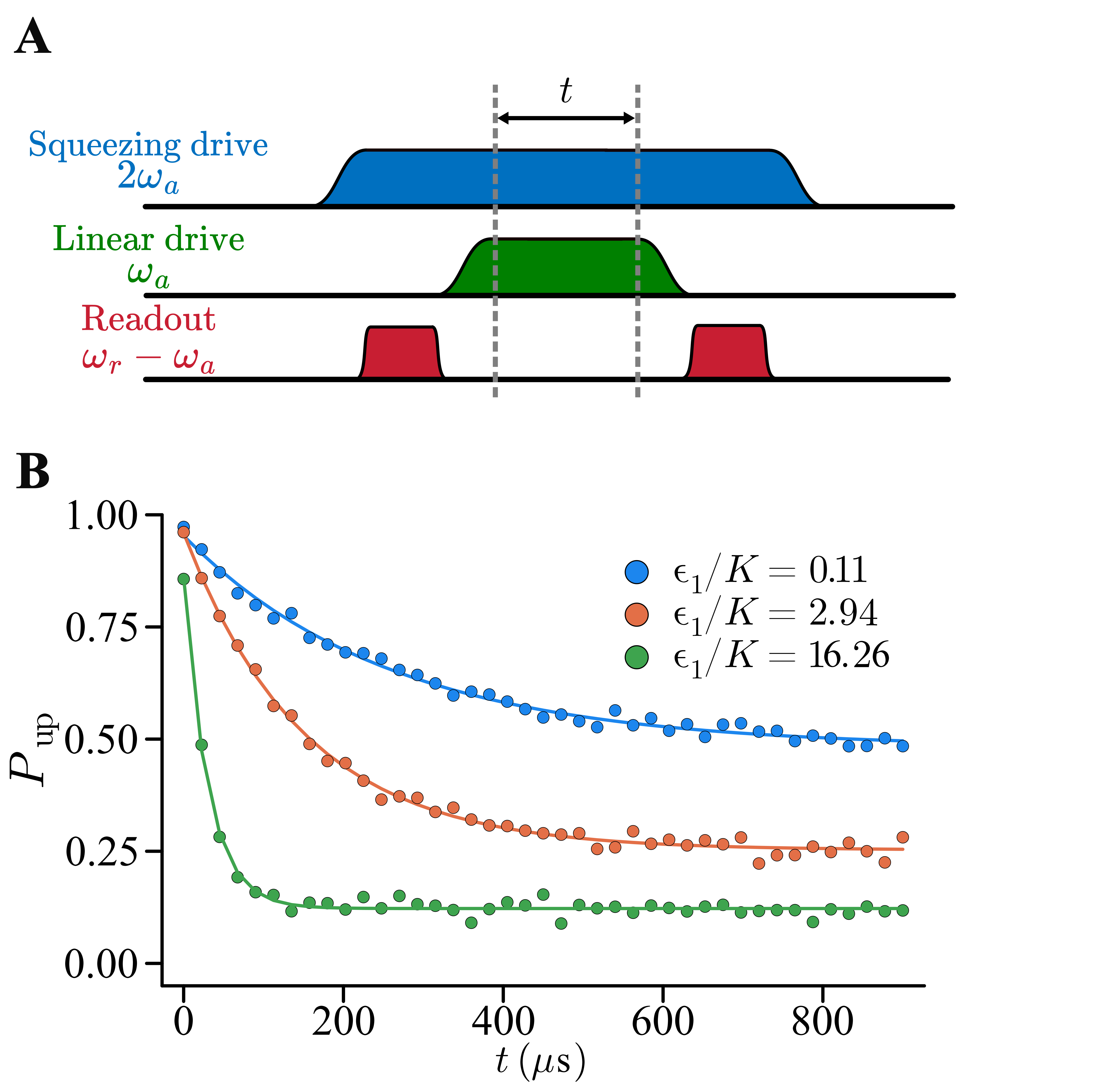}
    \caption{\textbf{Pulse sequence and typical decay curve for determination of activation time.} \textbf{A)} Pulse sequence for the determination of the activation time. The squeezing drive is turned on adiabatically. This is followed by a measurement of the which-well information, projecting the parametric oscillator into either of the wells. Then the linear drive is turned on adiabatically. Next, the state evolves for a variable time $t$, during which both the squeezing and linear drive remain on. After that, the linear drive is turned off adiabatically. Finally, the which-well information is measured again to find the remaining population. \textbf{B)} Decay of a coherent state initiated on the shallower well for different asymmetry values $\epsilon_1/K$. Experimental data are dots, solid lines are exponential fits. For small well asymmetry $\epsilon_1/K$, the probability to be in the shallower well decays to $0.5$. For increasingly large asymmetry, the steady-state population is no longer equally distributed between both wells, becoming increasingly biased towards the deeper well.}
    \label{fig:pulseSeq+exps}
\end{figure}

\subsection{Complementary analysis}
\subsubsection{Parameter extraction} \label{ParamExtract}
\begin{figure}[t!]
    \centering
    \includegraphics[width=\columnwidth]{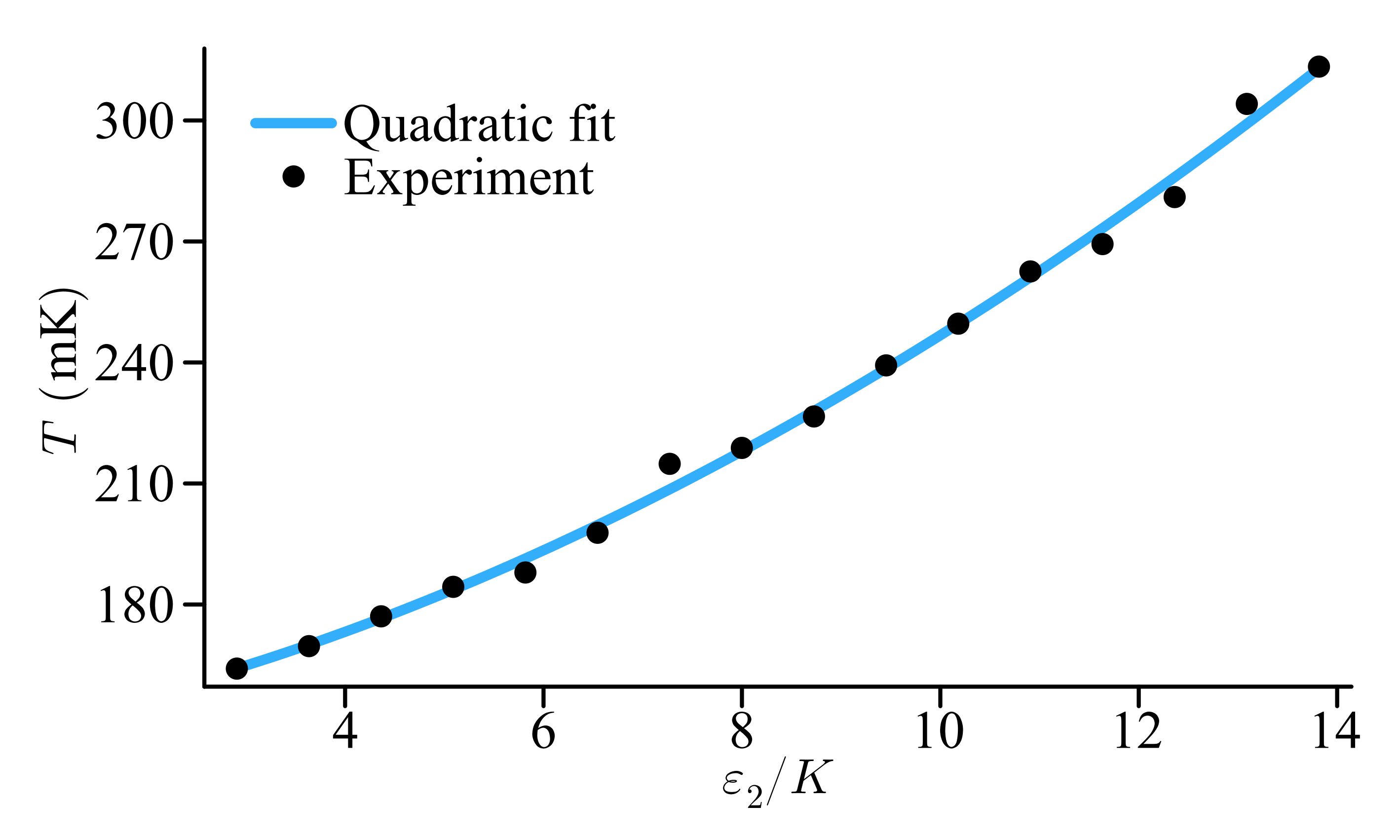}
    \caption{Temperature dependence on $\epsilon_2$. Round markers were extracted from experimental data in \cref{fig:mamba} \textbf{A} by fitting to theory.}
    \label{fig: temp vs ep2}
\end{figure}

Theory parameters $K$, $\Delta$, $\kappa$ and $n_\text{th}$ are obtained by fitting the Lindbladian model 
\begin{equation}
\label{eq:lindabladian}
    \dot{\hat{\rho}} = -\frac{i}{\hbar}[\hat{H}_{\text{eff}}, \hat{\rho}] + \kappa(1+n_\text{th})\mathcal{D}[\hat{a}](\hat{\rho}) + \kappa n_\text{th}\mathcal{D}[\hat{a}^\dagger](\hat{\rho})
\end{equation}
to the experimental data in \cref{fig:mamba} \textbf{A}, starting from the independently measured values. Here $\hat{H}_{\text{eff}}$ is the Hamiltonian from \cref{eq:H_eff} (with $\phi=0$). First, Kerr is found by fitting the location of the resonance parabolas in the $\epsilon_1$-$\epsilon_2$ plane of \cref{fig:mamba} \textbf{A} to those in \cref{fig:mamba} \textbf{B} since these features are very robust to changes in $\Delta$, $\kappa$ and $n_\text{th}$. This resulted in a Kerr of $K/2\pi = 465$ kHz. Next, for each value of $\epsilon_2$, $n_\text{th}$ is fitted so that the resulting timescale $\tau$ best quantitatively agrees with its experimental counterpart. With this, we report a quadratic dependence of the effective temperature $T$ on $\epsilon_2$, see markers in  \cref{fig: temp vs ep2}. We use a quadratic fit of $n_\text{th}$ on the theory, extending continuously those values of $\epsilon_2$ for which there is no experimental record of $n_\text{th}$, see solid line in \cref{fig: temp vs ep2}. Lastly, as reported in \cite{frattini_observation_2024, hajr_high-coherence_2024}, we fine-tuned $\Delta/2\pi = -558$ kHz and $\kappa/2\pi = 5.1$ kHz (corresponding to a single-photon lifetime of $31$ $\mu s$) 
to match the phase of the step-like oscillations of $\tau$ in the theory to those of the experiment. 

\subsubsection{Possible contributions to the effective temperature and mitigation techniques} \label{effTemp}
The extracted effective temperature has a quadratic dependence on $\epsilon_2/K$, as shown in \cref{fig: temp vs ep2}. In turn, $\epsilon_2/K$ is directly proportional to the voltage $V$ from the digital controller electronics (see \cref{fig:rabi plus nbar calib}). Therefore, $T \propto (\epsilon_2/K)^2 \propto V^2$, which is the form of Johnson-Nyquist noise. A natural explanation for this effective temperature dependence is thus a heating attenuator or 50 $\Omega$ termination (see \cref{fig: fridge} for the attenuator configuration). This could be mitigated by improved thermalization of attenuators, filters and terminations. Possible other contributions are the noise temperature of the room-temperature electronics, leakage via hybridization with the buffer mode (see \cref{fig: staircase and X } for some experimental evidence of that in the very strong drive regime), and unwanted higher-order loss channels in the SNAIL loop. Promising engineering solutions to these issues are on-chip filters \cite{hajr_high-coherence_2024}, careful detuning of the buffer mode \cite{benhayoune-khadraoui_how_2025}, and the Symmetrically Threaded SQUID as a replacement for the SNAIL \cite{bhandari_symmetrically_2024}. Regardless of the source of heating, engineered dissipation has been demonstrated for cooling of superconducting circuit KPOs \cite{ding_quantum_2025, adinolfi_enhancing_2025} and is an effective mitigation technique if lower effective temperatures are desired.

\subsubsection{Exponential decay} \label{expDec}
A crucial step in our data processing relies on fitting the activation dynamics with exponential decays. The experimental sequence is shown in \cref{fig:pulseSeq+exps} \textbf{A} and three exemplary decay curves for different values of the asymmetry are shown in \cref{fig:pulseSeq+exps} \textbf{B}. The data with the corresponding exponential fits is shown in \cref{fig:exponentials} \textbf{A}. We remark that each of the curves is well described by an exponential decay. Next, we notice that for small asymmetry $\epsilon_1 = 0.1$ the steady state population of the shallower well is around 50\%. This is because for a symmetric double-well the probability of being in either well is identical by construction, thus leading to equal steady-state populations. On the other hand, with increasing asymmetry, the tunneling rates from one well to the other become asymmetric, leading to a bias towards the deeper well. This is represented in our data, where for increasingly large asymmetries, the steady-state population of the shallower well is reduced.

\subsubsection{Control experiment for the relative phase dependence of the linear drive}
The relative phase $\phi$ between the squeezing drive and linear drive is determined by searching for the phase of maximum Rabi rate, which occurs at $\phi = 0$. The calibration measurement is shown in Fig. \ref{fig:cp03phase 180} \textbf{A}. This additional degree of freedom allows for several control experiments. To confirm that the resonant features we measure are actually caused by the controlled symmetry breaking, and are not for example just power-dependent non-linear resonances \cite{xiao_diagrammatic_2023, dumas_unified_2024} or due to Stark shifting into resonance with spurious modes \cite{thorbeck_readout-induced_2023, sivak_real-time_2023}, we measure the activation rate for a symmetric and an asymmetric double-well, both under an equally strong linear drives. To achieve this, we set the linear drive phase $\phi=90\degree$ and $\phi=180\degree$ respectively. The resulting measurement is shown in Fig. \ref{fig:cp03phase 180} \textbf{B} and \textbf{C} (see also Fig. \ref{fig: monster appendix} \textbf{E}). We note that the resonances are not present for $\phi=90\degree$, thereby proving that they are indeed a controlled effect from the symmetry breaking in the parametric oscillator. 

As discussed in the Sec. \ref{Exper}, we have full control over the Rabi phase (see \cref{fig:cp03phase 180} \textbf{A}) which controls how the symmetric double-well is perturbed. A phase of $\phi = 0\degree$ applies a drive that lifts the ``left" well (see \cref{fig: monster appendix} \textbf{A}) and a phase of $\phi=90\degree$ does not break the symmetry between the wells (see \cref{fig: monster appendix} \textbf{D}). If a Rabi phase of $\phi=180\degree$ is applied, then the definitions of left and right well swaps, meaning the right well becomes the shallower well and the left well becomes the deeper well. We can then make a control experiment where we initialize the system in the shallower (now right) well and then measure the population of the left well as a function of time and asymmetry. The resulting data (Fig. \ref{fig:cp03phase 180} \textbf{C}) shows the same increase in tunneling rate with larger asymmetry $\epsilon_1/K$ and resonances at specific values. The observation of the same physical phenomena confirms our understanding of the system and validates our explanations of the observed effects.

\begin{figure}[t!]
    \centering
    \includegraphics[width=\columnwidth]{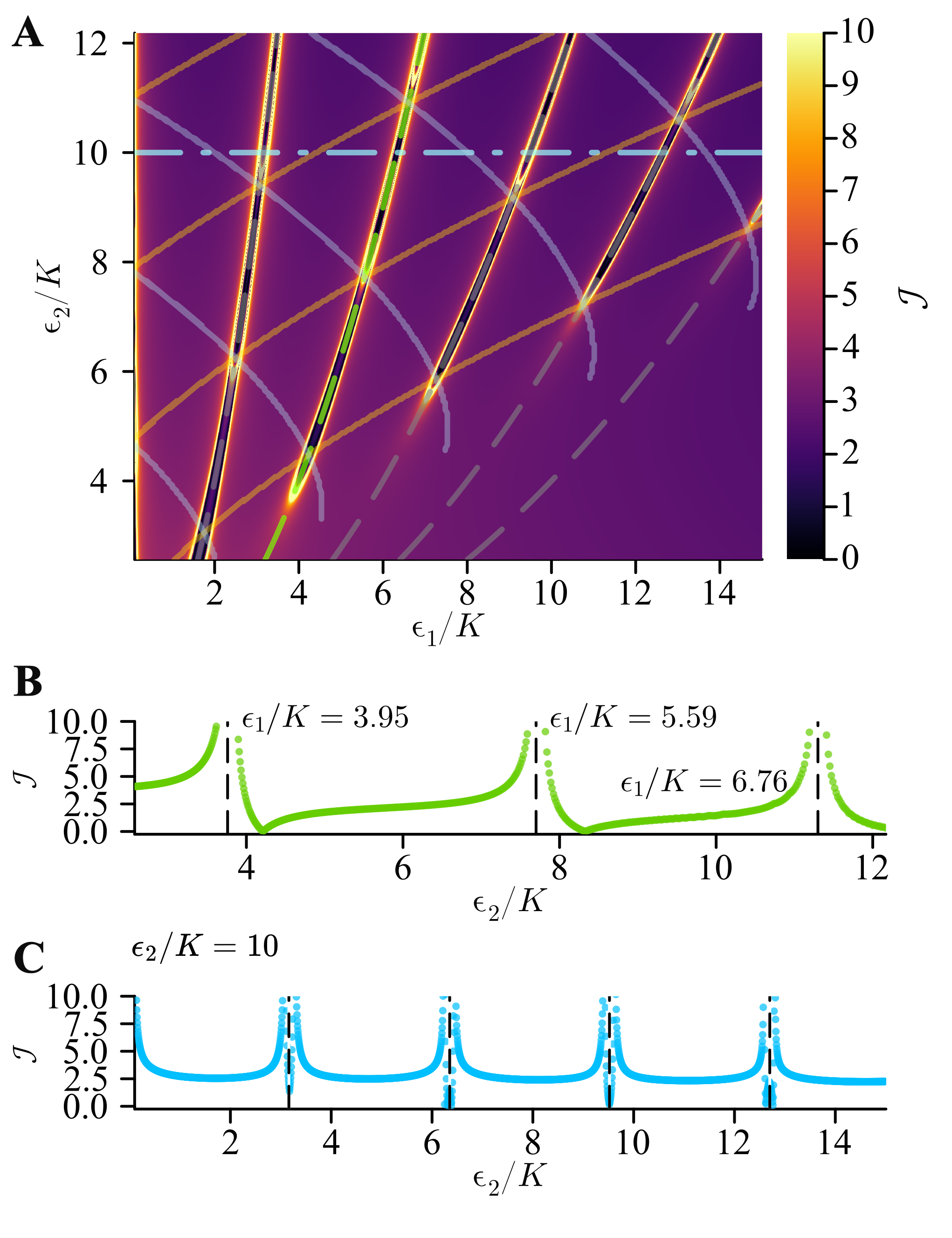}
    \caption{\textbf{Numerical quantum treatment and analysis.} \textbf{A)} Heatmap of $\mathcal{J}$ (numerical quantum treatment) with the semiclassical prediction on top (see \cref{fig:mamba} \textbf{C}). \textbf{B)} Plot of  $\mathcal{J}$ along the green dashed line in \textbf{A} parameterised by $\epsilon_2/K$. Vertical black dashed lines indicate the location of the singularities of $\mathcal{J}$ which happen very close to the triple blue-orange intersections. \textbf{C)} Plot of  $\mathcal{J}$ along the horizontal blue dashed line at $\epsilon_2/K=10$ in \textbf{A} parameterised by $\epsilon_1/K$. Vertical black dashed lines indicate the resonance condition from \cref{eq:parabola}. The different semiclassical predictions agree well with the full quantum Hamiltonian calculation.}
    \label{fig: ILAS}
\end{figure}

\begin{figure*}[t]
    \centering
    \includegraphics[width=\columnwidth*2]{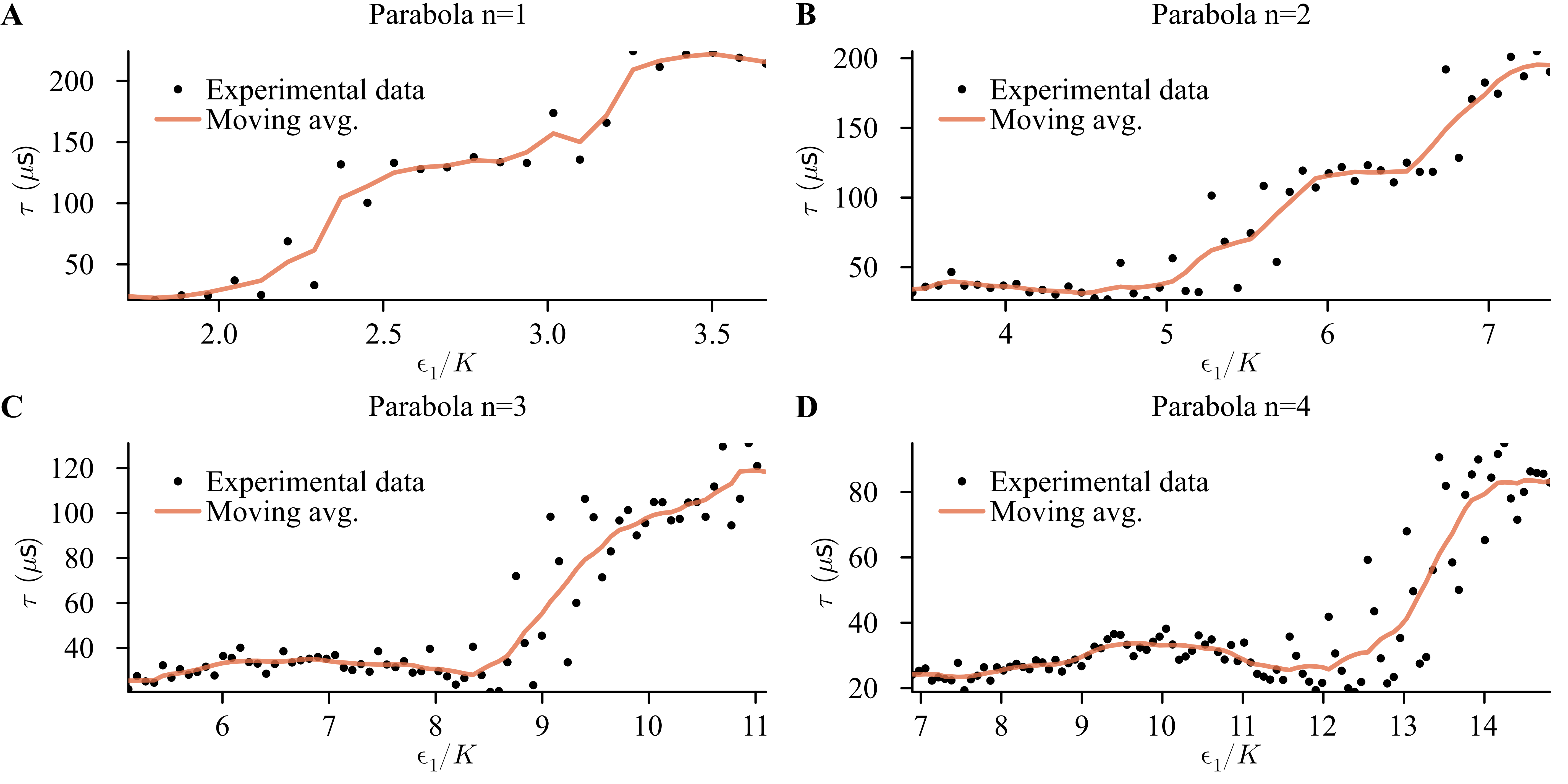}
    \caption{\textbf{Experimental lifetimes of \cref{fig:mamba} A along each resonance parabola as parameterised by $\epsilon_1/K$.} \textbf{A}-\textbf{D)} show the four parabolas for $n=1,2,3,4$ as defined by resonance condition \cref{eq:parabola}. The orange curves are guides to the eye (a moving average).}
    \label{fig:experimental staircases}
\end{figure*}

\subsubsection{Lifetime along resonances}
In \cref{fig:experimental staircases} we display the lifetime along the first four dashed parabolas of \cref{fig:mamba} \textbf{B}. As per resonance condition \cref{eq:parabola}, an increase in $\epsilon_1/K$ corresponds to an increase in $\epsilon_2/K$ when following the parabolas. Therefore, we expect the lifetime to increase along the parabolas, which we observe in \cref{fig:experimental staircases}. The step like behavior is a consequence of activation assisted tunneling close to the separatrix between the wells (see \cite{frattini_observation_2024}).

\subsubsection{Semi-classical analysis} \label{HamAna}
In \cref{fig: astral plus dbwells}, we provide a succinct pictorial view of the semi-classical mechanisms behind the features in \cref{fig:mamba} as described in the main text. 
\begin{figure}[t!]
    \centering
    \includegraphics[width=\columnwidth]{figs/astral_galaxy_plus_dwells.png}
    \caption{Semi-classical and Hamiltonian description of \cref{fig:mamba} \textbf{A}. \textbf{A)} Reproduction of \cref{fig:mamba} \textbf{C} with red circles highlighting the incoming resonant levels. \textbf{B)} Same as \textbf{A)} with cartoon double well potentials of \cref{eq:H_eff} with $H_\text{eff}$'s eigen energies drawn in the well corresponding to the location of its eigenstate. Note the resonance between left and right eigenstates and that the new incoming levels are near the top of the barrier marking the start of a new step in $\tau$. Double wells corresponding to dark red circles are not shown.}
    \label{fig: astral plus dbwells}
\end{figure}

\subsection{Full quantum Hamiltonian treatment}
In this section, we show that the physics captured by the semiclassical analysis used to explain the data is identically captured by a full numerical quantum treatment. For this, we introduce the inverse logarithmic anti-cross space (ILAS), denoted by $\mathcal{J}$ and defined as 
\begin{equation}\label{eq: ILAS}
    \mathcal{J} \doteq \sum_{n=0}^{N_\mathrm{cff}} \Bigg| \frac{1}{\log \left( E_{n+1} -E_n \right)}\Bigg|,
\end{equation}
where $E_n$ are the eigenenergies of \cref{eq:H_eff} at a particular $\epsilon_1/K$ and $\epsilon_2/K$ and $N_{\mathrm{cff}}$ is a numerical cutoff. We plotted a $\mathcal{J}$ colour-map as a function of $\epsilon_1/K$ and $\epsilon_2/K$, along with two different cuts in Fig. \ref{fig: ILAS}.

By construction, $\mathcal{J}$ diverges when a new pair of levels in $\{E_0,E_1,\cdots,E_{N_\mathrm{cff}}\}$ are closing. To see this, consider the scenario where the gap between the pair $E_0$ and $E_1$ and the gap between the pair $E_2$ and $E_3$ are closed, and the rest of the levels are not anti-crossing. In this situation, $\mathcal{J}$ is finite, since $\big|\frac{1}{\log \left( E_{n+1} -E_n \right)}\big|$ is zero for the two closed pairs and finite for $E_2-E_1$ and the rest of the levels (which are distant from one another). Consider now levels $E_4$ and $E_5$ interacting, starting to close their gap as in Fig. \ref{fig: wells plus spectrums} \textbf{D}. These interacting levels will close and at some point have an energy difference $E_4 - E_5 = 1$ where $\big|\frac{1}{\log \left( E_{n+1} -E_n \right)}\big|$ will diverge. In this way, $\mathcal{J}$ picks up poles near new anti-crossings reproducing \cref{fig:mamba} as can be seen in \cref{fig: ILAS} \textbf{B} and \textbf{C} in greater detail. Observe the close agreement between the semiclassical treatment, the resonance condition \cref{eq:parabola}, and $\mathcal{J}$.

\subsection{Effects beyond the RWA model}

\begin{figure}[ht]
    \centering
\includegraphics[width=\columnwidth]{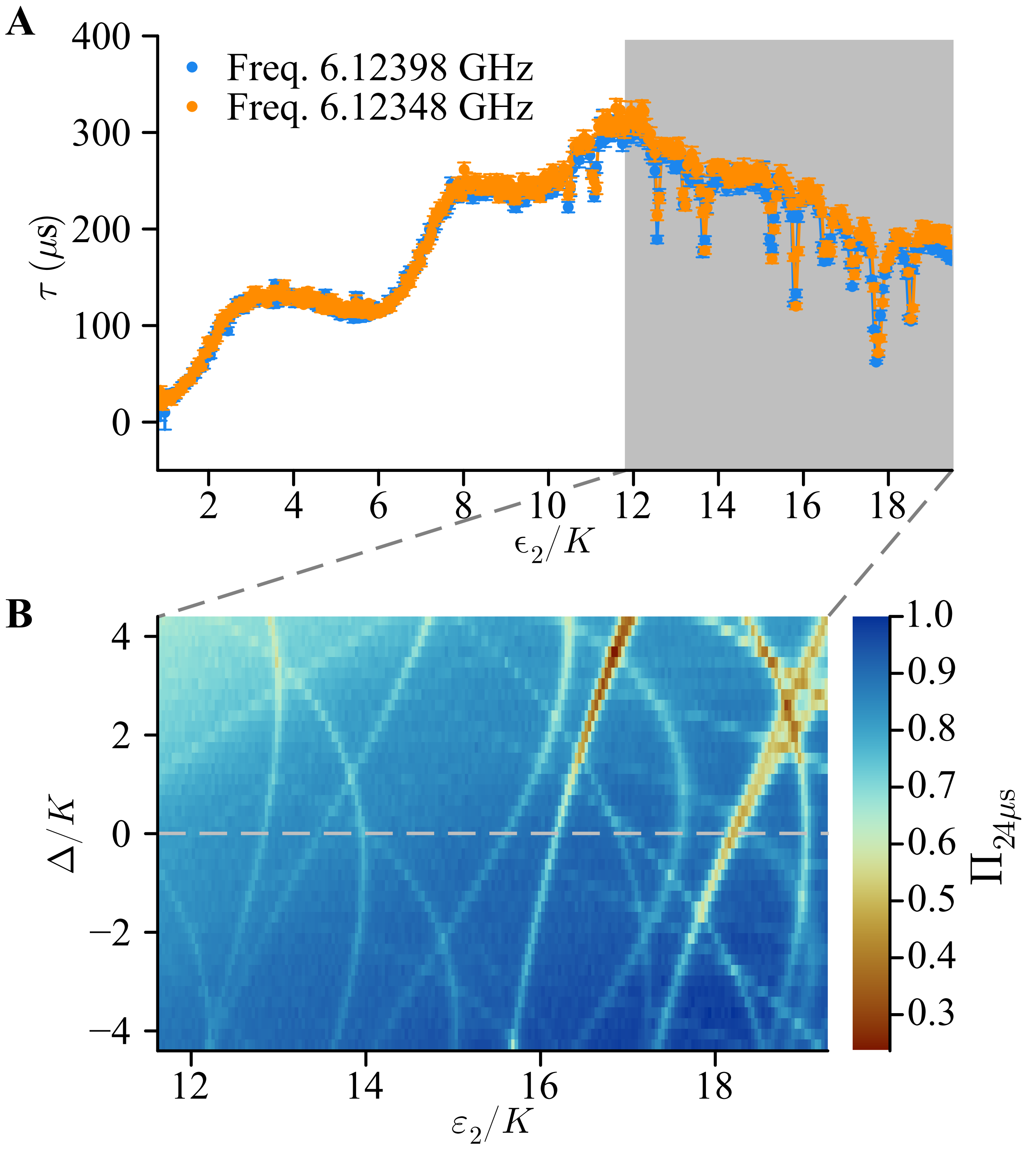}
    \caption{\textbf{Resonance-like features not explained by RWA model.} \textbf{A)} Coherent state lifetime as a function of photon number $\epsilon_2/K$ (at $\epsilon_1 = 0$). The blue and red curves are the same measurement repeated after a 24 hours time difference. The frequency was recalibrated to account for a small drift. The lifetime saturates at around 300~\textmu s. For large photon numbers $\epsilon_2/K$, resonance-like features appear. The location of these is stable within the 24 hours time difference. \textbf{B)} The parametric oscillator is initialized in one of the wells and after 24~\textmu s the remaining population in this well is measured. This is a proxy for the coherent state lifetime and thus allows us to identify resonances like in \textbf{A}, which emerge as a low remaining population. This is measured as a function of photon number $\epsilon_2/K$ and detuning $\Delta = \frac{\omega_2}{2} - \omega_a$ between the first subharmonic of the squeezing drive and the SNAIL transmon resonance frequency. At $\epsilon_2/K = 15$ and $\Delta/K = -2$ a crossing between resonances is visible. Another feature occurs at $\epsilon_2/K = 19$ and $\Delta/K = 2$, where the line of a resonance splits into two.}
    \label{fig: staircase and X }
\end{figure}

\begin{figure*}[t]
    \centering
    \includegraphics[width=2\columnwidth]{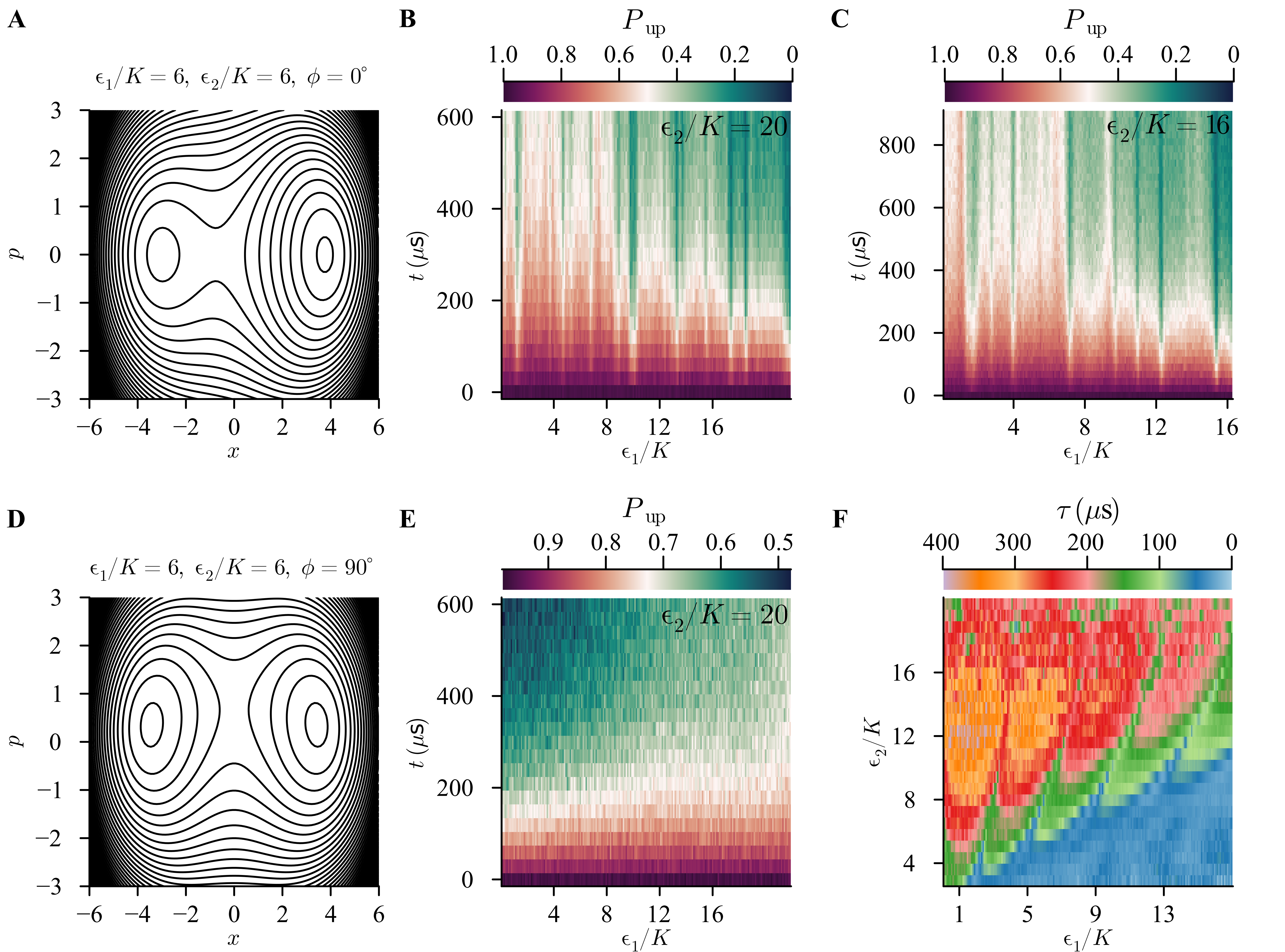}
    \caption{\textbf{Emergence of resonance-like features depending on different parameters.} \textbf{A}-\textbf{C)} Relative phase $\phi = 0\degree$ between linear drive and squeezing drive. \textbf{A)} Equienergy contours of the classical RWA parametric oscillator Hamiltonian with linear drive. The linear drive raises the left well and breaks the symmetry between the two wells within the RWA. \textbf{B)} Same experiment as in Fig. \ref{fig:exponentials} \textbf{A}, now at $\epsilon_2/K = 20$. A ``forest" of resonances is now visible besides the RWA resonances from condition \cref{eq:parabola}. \textbf{C)} Changing the photon number to $\epsilon_2/K = 16$ changes the resonance pattern. Note the different widths. \textbf{D}-\textbf{E)} Relative phase between linear and squeezing drive of $\phi = 90\degree$. \textbf{D)} Equienergy surfaces in the presence of the linear drive within the RWA, showing that the symmetry between wells is preserved and a small deformation in the $p$ direction is visible. \textbf{E)} Same measurement as in \textbf{B} but with $90 \degree$ shifted phase $\phi$. This change in phase is sufficient to remove all of the resonances, while the amplitude of both drives remains high. \textbf{F)} Extension of data in Fig. \ref{fig:mamba} \textbf{B}. Now the squeezing amplitude extends beyond $\epsilon_2/K = 14$ up to $\epsilon_2/K = 19$. The data in \textbf{B}, \textbf{C} shows up as flecks of short lifetime (blue) that are not captured by the RWA.}
    \label{fig: monster appendix}
\end{figure*}

In this section, we present experimental data taken for $\epsilon_2/K>14$ where the static effective (RWA) description used fails, qualitatively, to describe experiment. In Fig. \ref{fig: staircase and X } \textbf{A}, we present the coherent state lifetime up to $\epsilon_2/K \approx 19$, for $\epsilon_1=0$. The lifetime increases in a step-like fashion, as predicted by the static effective model \cite{putterman_stabilizing_2022,gautier_combined_2022} until it starts decreasing. This decrease may be captured by a more elaborated static effective treatment \cite{venkatraman_static_2022-1}. However, resonant-like drops of the lifetime appear too, which challenge the static effective treatment, even if they can also be captured in principle \cite{xiao_diagrammatic_2023}. Time-dependent Floquet simulation of the driven system (not shown, see \cite{garcia-mata_effective_2024}) suggest that these resonances may be explained by a single-mode treatment of the nonlinear transmon oscillator. Note that, since these experiments were conducted with $\epsilon_1 = 0$, these resonances are of a similar nature to the ones presented in the main text. Below, we present and discuss control experiments supporting the viewpoint that these effects belong to the nonlinear physics of single-mode driven systems.

As a control experiment to learn more about these resonances, we measure their temporal stability by repeating the same measurement again after 24 hours (see orange and blue graphs in Fig. \ref{fig: staircase and X } \textbf{A}). This provides important information since the parameters of spurious two-level-systems (TLS) \cite{klimov_fluctuations_2018, chen_phonon_2024, thorbeck_two-level-system_2023} coupled to transmons are known to fluctuate in this timescale, and TLS are a plausible candidate for this resonant-like lifetime drops at strong drives.
We re-calibrated the qubit frequency and observed good agreement between the two measurements, with the resonances being located at the same positions. This data suggests that the resonances are stable in time and are therefore likely associated to stable electromagnetic transitions and not fluctuating TLSs.

Next, we study these resonances as a function of the detuning $\Delta = \frac{\omega_2}{2} - \omega_a$ between the squeezing drive and the resonance frequency of the SNAIL transmon. The system is initialized in one well and after 24~\textmu s the remaining population is measured. This is a proxy for the coherent states' lifetime and thus resonances manifest themselves as a low remaining population. Fig. \ref{fig: staircase and X } \textbf{B} depicts this lifetime proxy as function of detuning and photon number. This data contains resonances moving in different directions. Other features can be seen, for example, around $\epsilon_2/K = 15$ and $\Delta/K = -2$: here two resonances cross (i.e., no avoided crossing), which could provide a bound to the possible coupling between the involved modes. Lastly, we point out a feature around $\epsilon_2/K = 19$ and $\Delta/K = 2$, where one of the resonances seems to split up into two. 

Moving beyond the case of $\epsilon_1 = 0$, we now study the impact of an additional drive on the resonances. In Fig. \ref{fig: monster appendix} \textbf{A}, we show the equienergy contours of the effective Hamiltonian for $\epsilon_1 \neq 0$ and $\phi = 0$, which creates an asymmetry between the two wells (see also Fig. \ref{fig:wells} \textbf{C}). This picture is valid under the RWA, and we have observed its conspicuous failure for $\epsilon_2 / K \gg1$. In \cref{fig: monster appendix} \textbf{B}, for $\epsilon_2/K = 20$ we observe a ``forest" of resonances, very different from the case of smaller $\epsilon_2/K$ (see Fig. \ref{fig:exponentials}). These resonances are not explained by resonant tunneling in the RWA potential and described by \cref{eq:parabola}, but are instead reminiscent to the onset of chaotic behavior \cite{chavez-carlos_driving_2024, garcia-mata_effective_2024,garcia-mata_chaos_2024}. Changing the photon number (see Fig. \ref{fig: monster appendix} \textbf{C}) leads to a change in the forest of resonances. 
Similar resonances have been observed during the readout of transmon qubits. Here a possible explanation is that the AC Stark shift, induced by the strong readout drive, tunes the qubit into resonance with lossy modes like TLS \cite{sivak_real-time_2023, sank_measurement-induced_2016, thorbeck_readout-induced_2023}. To test this hypothesis, we change the phase $\phi$ of our strong linear drive. For $\phi = 90 \degree$, the RWA double-well stays symmetric (Fig. \ref{fig: monster appendix} \textbf{D}). In \cref{fig: monster appendix} \textbf{E}, we show the same measurement as in Fig. \ref{fig: monster appendix} \textbf{B}, (same value of $\epsilon_1$), but with a different relative phase between the drives. The absence of resonances in Fig. \ref{fig: monster appendix} \textbf{E} suggests that the effect is beyond a Stark shift into lossy modes. This phase dependence of the spurious resonances has not been observed before.
Finally, \cref{fig: monster appendix} \textbf{F} shows an extended data set (compare to Fig. \ref{fig:mamba}), where the breakdown of the RWA description is self-evident for $\epsilon_2/K>14$.

While these spurious resonances have not been reported before, the discrepancy in between static-effective open quantum system description of our strongly driven nonlinear system and experimental observations is common to all parametric oscillator experiments in the quantum regime reported in the literature \cite{grimm_stabilization_2020,frattini_observation_2024,venkatraman_driven_2023,wang_quantum_2019,iyama_observation_2024,hajr_high-coherence_2024,yamaguchi_spectroscopy_2024,hajr_high-coherence_2024}, etc. We believe this discrepancy runs deep into our understanding of quantum physics \cite{noauthor_einsteins_nodate} and is intimately related to the problem of the quantum to classical transition and quantum chaos \cite{chavez-carlos_driving_2024}. We also believe it can be avoided once it is understood, by means as simple as filtering the lines at the relevant frequencies, for example.

Our data set, presenting resonances at different locations in parameter space and of different widths, as well as their dependence on the parametric drive frequency, the parametric drive amplitude, the linear drive amplitude, the relative phase between the linear drive and the parametric drive, and their stability over a period of 24~hs, will guide research and lead to better theoretical tools to understand and design parametric processes. We also expect that this data will unlock new tools and proposals to study nonlinear driven quantum systems and quantum chaos in unexplored regimes \cite{habib_decoherence_1998, zurek_decoherence_1994,zurek_why_nodate, garcia-mata_chaos_2024}.

\clearpage
\bibliography{AsymDoubleWell}

\end{document}